\documentclass[12pt, oneside]{article} 

\usepackage{amsmath}  				
\usepackage{geometry}
\usepackage{amssymb}
\usepackage{mathtools}
\usepackage{slashed}
\usepackage[parfill]{parskip}    		
\usepackage{graphicx}
\usepackage{mathtools}

\allowdisplaybreaks
\usepackage[justification=centering]{caption}
\usepackage{mathtools}
\numberwithin{equation}{section}

\usepackage{xfrac}
\usepackage{amssymb}
\usepackage[table]{xcolor}
\usepackage{booktabs}
\usepackage{mdwtab}
\usepackage{breqn}
\usepackage{amsthm}
\usepackage{lmodern}
\def\a{\alpha}
\def\b{\beta}

\def\d{\delta}
\def\e{\epsilon}

\def\g{\gamma}

\def\m{\mu}
\def\n{\nu}
\def\o{\omega}

\def\r{\rho}
\def\s{\sigma}

\def\F{\Phi}
\def\G{\Gamma}

\usepackage{float}
\usepackage{xcolor}
\usepackage{lscape}
\usepackage{subfig}
\usepackage{amsthm}
\usepackage{feynmp-auto}
\usepackage{graphicx}
\usepackage{caption}
\usepackage{subcaption}

\def \nn {\nonumber}
\def \be  {\begin{equation}}
	\def \ee  {\end{equation}}
\def \bea  {\begin{eqnarray}}
	\def \eea  {\end{eqnarray}}

\begin{document}
	\begin{flushright}
		BONN-TH-2025-17
	\end{flushright}
	\begin{center}
		\Large \bf Power corrections to the heavy electron form factor
	\end{center}
	
	\bigskip
	
	\centerline{ Aniruddha Venkata$^a$}
	\begin{center}
		${}^a$Bethe Center for Theoretical Physics, Universit$\ddot{\text{a}}$t Bonn, D-53115, Germany
	\end{center}
	\date{}
	\abstract{We study the first power correction to the heavy electron form factor in QED and show that it factorizes as a derivative operator. We discuss the result in QED with no light fermions, where the first power correction can be written explicitly in terms of one-loop integrals and the anomalous magnetic moment. In the presence of light fermions, the heavy electron form factor admits a representation as a sum over matrix elements, each of which receives corrections from higher orders in perturbation theory. From this analysis, we are able to extract the next-to-leading power soft photon theorem in the limit of  heavy fermion-initiated dijet events. This is a first step towards studying the heavy quark form factor in the non-abelian theory.  }

	\section{Introduction}
	\label{sec:intro}
	
The investigation of power corrections to QCD observables is of importance in the era of precision collider physics \cite{Boughezal:2022cbl},\cite{Maltoni:2022bqs},\cite{Heinrich:2020ybq}. Studying power corrections to jet observables at the cross-section level has been an area of active research and has attracted increasing interest \cite{Caola:2021kzt},\cite{vanBeekveld:2021hhv},\cite{Liu:2020tzd}, \cite{Beneke:2019mua}, and \cite{Mannel:2010wj}.  
 Such corrections to N-subjettiness \cite{Thaler:2010tr}, a jet shape observable designed to identify hadronically decaying heavy particles, have been studied extensively \cite{Moult:2016fqy},\cite{Boughezal:2016zws}, \cite{Ebert:2018lzn}, \cite{Boughezal:2019ggi}. Therefore, the  study of power corrections to observables involving the production and decay of heavy quarks is of interest to the collider physics community. 
 
 A complementary approach useful for resummation is studying power corrections at the level of amplitudes or form factors \cite{Laenen:2020nrt},\cite{vanBijleveld:2025ekz}, \cite{terHoeve:2023ehm}. In these works, the form factor was investigated in QED in the light fermion limit. However, in this limit, the form factor consists of four independent components: two jet functions, a soft function, and a hard function. Schematically, one may write 
 \bea
 \Gamma(p_1,p_2)=J_{1}(p_1,\beta_2)H(p_1,p_2)J_{2}(p_2,\beta_1)S(\b_1,\b_2), \label{eq:factorized-form}
 \eea
 where $\b_{i}=\frac{p_i}{p_i^0}$. The functions $J_1,$ and $J_2$ are the jet functions, the function $S$ is the soft function, and $H$ is the hard function.  In studying power corrections to the form factor in the light fermion limit, all four functions contribute to the next-to-leading power (NLP) corrections. Further, as a result of pinch surfaces with physically polarized partons connecting hard and jet functions, which begin to contribute at NLP, the simple factorized form in Eq.\ (\ref{eq:factorized-form}) is modified into a more complicated convolution between hard and jet modes. This makes the study of factorization theorems formidable even in QED. Such an NLP study of the form factor in the light quark limit is useful to study the thrust cross section in the two-jet limit and is of importance phenomenologically. 
 
 At the level of a cross-section, there are multiple notions of a power correction since  cross-section predictions are often at leading power in multiple dimensionless parameters. For example, the thrust cross-section in $e^{+}e^{-}$ annihilation \cite{Sterman:1995fz}, depends on the dimensionless ratio $\frac{Q^2}{\mu^2}$ (here $Q^2$ is the centre of mass energy and $\mu$ is the renormalization scale) as well as the thrust variable $1-T$. Therefore, the thrust cross-section admits a power expansion in two parameters $1-T,\frac{\Lambda_{QCD}}{Q}$.  At leading power in $1-T$, the thrust cross-section develops a large logarithm of the form $\sim \log^2 (1-T)$, which corresponds to a singularity in the two-jet limit. Leading power factorization of the Sudakov form factor in Eq. (\ref{eq:factorized-form}) allows us to resum all large logarithms in $1-T$. In contrast, power corrections to the total cross-section in  $e^{+}e^{-}$ annihilation are corrections of the form $\frac{\Lambda_{QCD}^4}{Q^4}$ and arise from vacuum condensates \cite{Shifman:2010zzb}. In what follows, our interest is in power corrections of $O\left(\frac{\Lambda_{QCD}}{Q}\right)$.  
 
\begin{figure}[h]
 	\begin{center}
 		\begin{tabular}{cc}
 			\includegraphics[width=.45\textwidth]{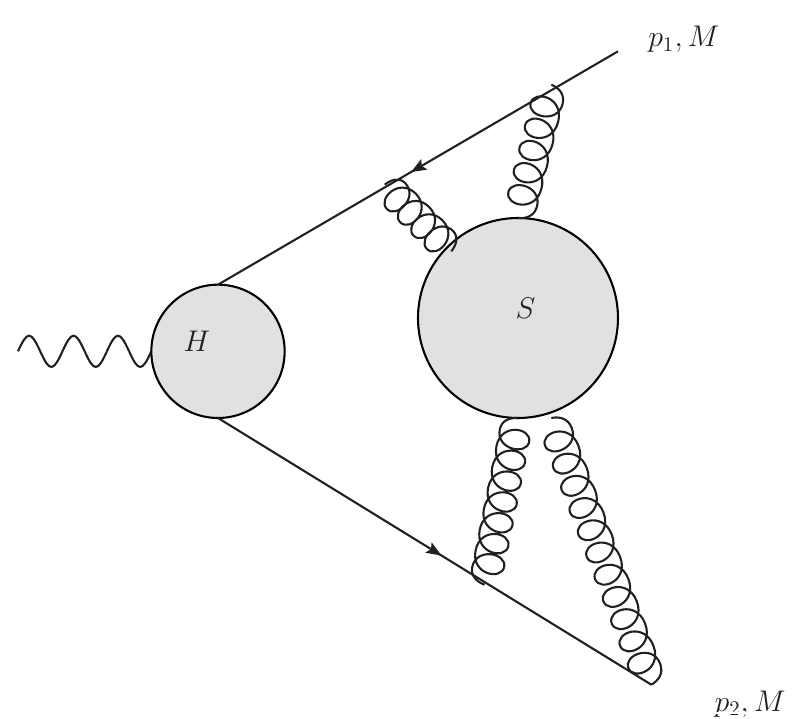}\hfill &
 			\includegraphics[width=.45\textwidth]{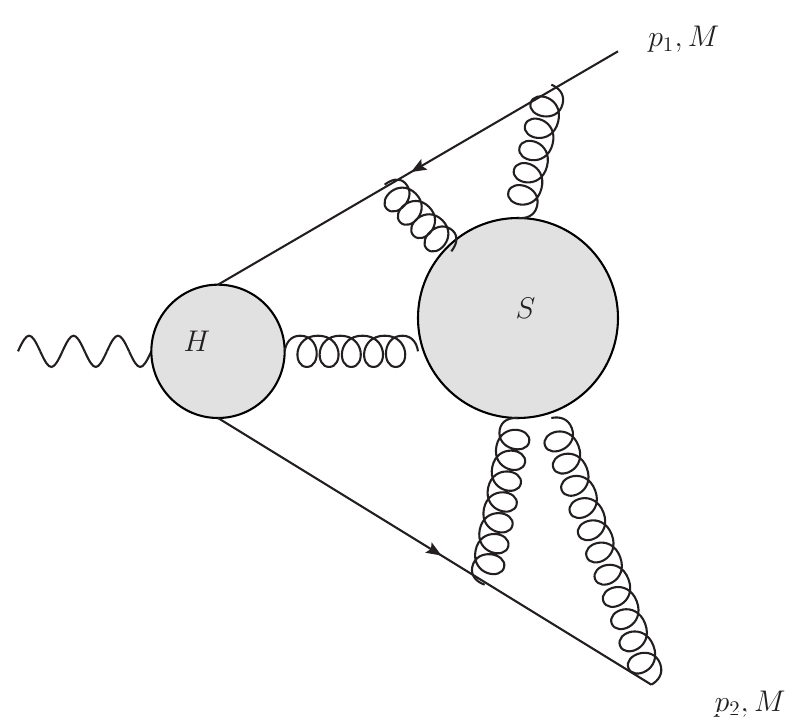}\hfill \\ 
 			(a) &(b)
 		\end{tabular}
 		\caption{Two types of pinch singular surfaces that are present in the  form factor.  (a) Leading pinches (b) Non-leading pinches } 
 		\label{fig:pinches}
 	\end{center}
\end{figure}

  Another simple class of questions involves power corrections to the form factor in the  heavy quark limit. The heavy quark form factor which we denote $\Gamma^{\r}(p_1,p_2,M)$ at leading power reads \cite{Korchemsky:1991zp}, 
 \bea
 \Gamma^{\r}(p_1,p_2,M)=H^{\r}(p_1,p_2,M) \langle W_{v_1}(0,\infty)W_{v_2}(0,\infty)\rangle. \label{eq:heavy-quark-LP}
 \eea 
Here, $v_i^{\mu}$ are the $3$-velocities in the $p_i$ direction and $W_{v_i}$ are Wilson lines in th $v_i$ direction.  In particular, the velocities $v_i$ are four vectors defined to have unit energy.
 The Wilson lines take the explicit form
 \bea
 W_{v_1}(0,\infty)&=&\mathcal{P}\left(e^{ig\int_{-\infty}^{0}d\s\,  v_1^\m A_{\m}(v_1\s)}\right)\nn\\
  W_{v_2}(0,\infty)&=&\mathcal{P}\left(e^{ig\int_{0}^{\infty}d\s\,  v_2^\m A_{\m}(v_2\s)}\right). 
 \eea
 The perturbative expansion of the Wilson lines in momentum space requires regularization to avoid eikonal singularities. This is completely determined by the limits of $\s$ integration and demanding that integrals are well defined at $\infty$.  
  It will be convenient  to study the form factor with incoming momentum $q=p_1+p_2$, in the centre of mass frame
 \bea
 p_1&=&\left(\sqrt{p^2+M^2},0_{\perp},p\right)\nn \\
 p_2&=&\left(\sqrt{p^2+M^2},0_{\perp},-p\right)\nn\\
 q&=&\left(2\sqrt{p^2+M^2},\vec{0}\right).
 \eea
 In this frame, the velocity vectors can be made explicit
 \bea
 v_{1}=\left(1,0_{\perp},\frac{p}{\sqrt{p^2+M^2}}\right)\nn \\
 v_{2}=\left(1,0_{\perp},\frac{-p}{\sqrt{p^2+M^2}}\right). 
 \eea
 Further, we will use Feynman gauge, where the gluon propagator is 
 \bea
 \langle A_{\mu}^a(k)A_{\nu}^b(k')\rangle=\frac{-i\d^{ab}\eta_{\m\n}}{k^2+i\e}(2\pi)^d \d^d\left(k+k' \right).
 \eea
 With these conventions in place, we can explain the formula in Eq.\ (\ref{eq:heavy-quark-LP}). The function $H(p_1,p_2,M)$ is a IR finite, hard function which we may evaluate in perturbation theory, while the Wilson line expectation value is an IR divergent soft function which involves the exchange of soft modes. The absence of an IR divergent jet function is explained by the presence of massive quarks in the final state, with a mass scale $M \gg \Lambda_{QCD}$. The pinch singular surfaces of the form factor are shown in Fig.\ \ref{fig:pinches}. 
 
 The leading pinches in Fig.\ \ref{fig:pinches} have no soft gluons connecting the soft function and the hard function, while the sub-leading pinches in Fig.\ \ref{fig:pinches} do. Loop integration of the form factor in the vicinity of a soft pinch involves integration of modes that satisfy $|\vec{k}|< \Lambda_{QCD}$ and, by definition, are integrals over soft scales, yielding results of $O(\Lambda_{QCD})$. At leading power, integration of loop momentum modes satisfying $|\vec{k}|< \Lambda_{QCD}$ can be separated out, and the resulting expression is Eq.\ (\ref{eq:heavy-quark-LP}). Qualitatively, there are two sources of corrections to the formula in Eq.\ (\ref{eq:heavy-quark-LP}): integration around sub-leading soft pinch surfaces and subleading terms near leading soft pinch surfaces. To arrive at a formula analogous to Eq.\ (\ref{eq:heavy-quark-LP}) at NLP, we must account for both types of corrections and separate out soft modes from hard modes. The presence of subleading pinches that connect the hard part and the soft subgraph suggests that at NLP, there is a convolution between the hard and soft functions. We will show that in QED, this expectation is wrong, and conventional factorization extends to NLP, with new soft functions. 
 
 Recently, it was shown that the soft photon theorem \cite{Low:1958sn},\cite{Burnett:1967km},\cite{Weinberg:1965nx}, \cite{DelDuca:1990gz} at leading power is modified in the presence of massless quarks \cite{Ma:2023gir}. A detailed study at next-to-leading power in all  two-loop QED graphs has been carried out in \cite{Laenen:2020nrt}. A more comprehensive all-loop theorem is still missing in the massless quark limit. It is not likely that the radiative amplitude is related to the nonradiative form factor alone. It is therefore worthwhile to investigate the opposite, massive quark limit, and ask if the NLP soft photon theorem is modified in this limit as well. Developing such an NLP soft photon analysis requires NLP factorization, which is the subject of study here. We will also state the NLP soft photon theorem in the heavy electron limit in Sec.\ \ref{sec:all-order-analysis}. In QCD, an explicit cutoff can be placed around each pinch surface, which is a UV cutoff for the soft modes and an IR cut off for the hard function as previously described. Such a cutoff is physical when the form factor is embedded inside a cross-section (like a thrust cross-section). However, in carrying out explicit calculations, it is natural to consistently use dimensional regularization both in QCD and QED.   In this setting, the leading power soft function is defined by its ultraviolet (UV) counter-terms, and the leading power soft function contains UV poles accompanied by a factorization scale, $\m$. Power corrections to  this soft function are therefore naturally organized by the ratio $\frac{\m}{p_i\cdot v_i}$.  This structure is observed in our explicit results in Eqs.\ (\ref{eq:one-loop-NLP}),(\ref{eq:final-result}).

 In this work, we study the question of NLP corrections to the heavy quark form factor in the abelian theory. We will therefore study the heavy electron form factor in QED, with and without virtual light fermions in intermediate states. In Sec.\ \ref{sec:one-loop-analysis}, we study the one-loop form factor graphs in massive QED and arrive at an NLP factorization formula valid at one-loop. We will find that the one-loop factorization formula is gauge invariant and reproduces the one-loop triangle graph at NLP in a perturbative expansion of the matrix element. In Sec.\ \ref{sec:two-loop-analysis}, we repeat this analysis  at two loops and arrive at additional terms that begin at two loops. We find that the new terms are independently gauge invariant. In Sec.\  \ref{sec:all-order-analysis} we argue that there are no new terms  differing from those inferred in the two-loop analysis. In Sec.\  \ref{sec:conclusions}, we summarize our results and list directions for generalization that we plan to study in future work.
 \section{One-loop analysis}
 \label{sec:one-loop-analysis}
\begin{figure}
 	\centering
 	\includegraphics[width=0.45\textwidth]{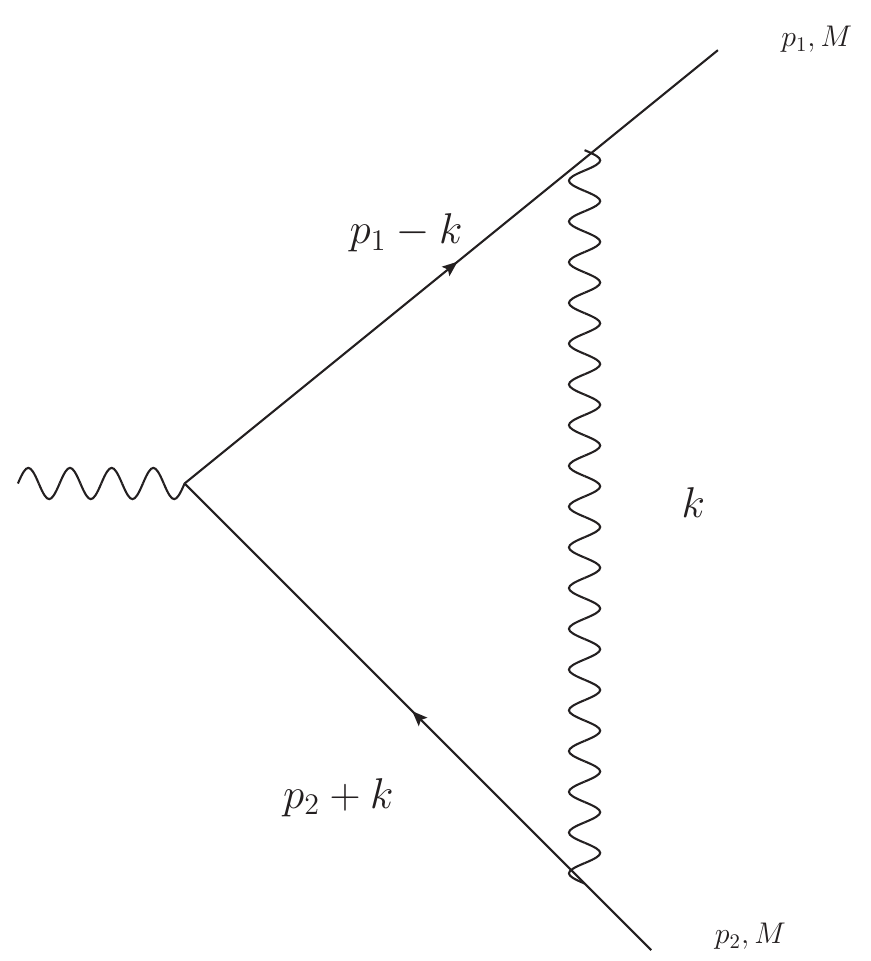}
 	\caption{The one-loop contribution to the form factor}
 	\label{fig:ladder}
 \end{figure}

 In this section, we study the one-loop soft function in QED. Let us first observe that in QED, the one-loop ladder graph is IR divergent in four dimensions at leading power. Further, it exponentiates and captures all IR divergences in QED \cite{Grammer:1973db}. The soft function can be evaluated at all loops by evaluating a one-loop integral
 \bea
 \langle W_{v_1}(0,\infty)W_{v_2}(0,\infty)\rangle=\exp \left(-ie^2\m^{2\e} \int\frac{d^{4-2\e}k}{(2\pi)^{4-2\e}} \frac{-v_1 \cdot v_2}{(-v_1\cdot k+i\e)(v_2 \cdot k+i\e)(k^2+i\e)} \right) \label{eq:exponentiation}
 \eea
 The relevant graph at one-loop is the QED ladder graph shown in Fig.\ \ref{fig:ladder}.
 The vertex correction at one-loop reads
 \bea
 \bar{u}(p_1,s_1)\tilde{\Gamma}^{\rho,\, (1)}(p_1,p_2,M)v(p_2,s_2)&=&\int\frac{d^{4-2\e}k}{(2\pi)^{4-2\e}}\frac{\bar{u}(p_1,s_1)(-ie\g^{\m})\left(i[\slashed{p_1}-\slashed{k}+M]\right)\g^{\rho}}{((p_1-k)^2-M^2+i\e)} \nn \\ &&\hspace{-1cm}\times \frac{ \left(i[-\slashed{p_2}-\slashed{k}+M]\right) (-ie\g^{\nu}) v(p_2,s_2) \left(-i\eta_{\m\n}\right)}{((p_2+k)^2-M^2+i\e)(k^2+i\e)}.\label{eq:one-loop-FF}
 \eea
 Here, we work in a normalization where  $\tilde{\Gamma}^{\rho,\, (0)}=\gamma^{\r}$. 
 This graph has only one pinch surface at $k^{\m}=0$, the soft pinch. Such a pinch is a leading pinch since the tree level hard function, $\tilde{\Gamma}^{\rho,\, (0)}$, is independent of $k$. At one-loop, there are no non-leading pinches. In anticipation of the distinction from the two-loop case discussed in Sec.~\ref{sec:two-loop-analysis}, we note that at two loops, both subleading and sub-subleading pinches can arise. Let us make an observation about the notation we use here. We use the symbol $\tilde{\Gamma}^{\rho}$ to denote the ladder graph contribution to the full form factor $\Gamma^{\rho}$. These differ by the insertion of counter-terms as well as self-energy graphs, which we don't analyze in this section.
 
  At leading power, in Eq.\ (\ref{eq:one-loop-FF}) we make the following approximations near the pinch
 \begin{itemize}
 	\item In the numerator, we drop $\slashed{k}$ in favour of $\slashed{p}_1+M$. Clearly, the term with $\slashed k$ in the numerator has an additional vanishing scale, which is power suppressed. 
 	\item In the $p_1$ denominator, we make the approximation $(p_1-k)^2-M^2+i\e \approx -2p_1 \cdot k+i\e$, dropping the $k^2$ relative to $-2p_1 \cdot k$. This follows from the soft approximation and is justified by the assumption of outgoing kinematics, as discussed in \cite{Collins:1988ig}. In this case, since both $p_1,p_2$ are outgoing, we may make this approximation safely.   
 	\item  In the $p_2$ denominator, we make the approximation $(p_2+k)^2-M^2+i\e \approx 2p_1 \cdot k+i\e$, dropping the $k^2$ relative to $2p_2 \cdot k$.
 	\end{itemize} 
 	
Having made a list of approximations to the one-loop graph, we may use the Dirac equation on either spinor, $\bar{u}(p_1,s_1)(\slashed{p}_1-M)=(\slashed{p}_2+M)v(p_2,s_2)=0$, to write the approximated integrand as
\bea
\tilde{\Gamma}^{\rho,\, (1)}_{\text{LP}}(p_1,p_2,M)=-ie^2\m^{2\e} \int\frac{d^{4-2\e}k}{(2\pi)^{4-2\e}} \frac{-v_1 \cdot v_2 \; \g^\r}{(-v_1\cdot k+i\e)(v_2 \cdot k+i\e)(k^2+i\e)}, \label{eq:one-loop-lp}
\eea
which agrees with the leading term in the expansion of Eq.\ (\ref{eq:exponentiation}) multiplying the tree-level hard function $\g^\r$. We can now separate the next-to-leading power terms in Eq.\ (\ref{eq:one-loop-FF}). Of the two  $\slashed{k}$ in the numerator, we may retain one and continue to make the eikonal approximation. Alternatively, we may expand the eikonal denominator and consider next-to-eikonal corrections. Next-to-eikonal corrections have been studied in non-abelian gauge theories in \cite{Laenen:2010uz}. Here, we specialize to the case of abelian gauge theory with fermionic matter. 

To isolate the contributions, we analyze each fermion line separately and define
\bea
\bar{u}(p_1,s_1)F_1^{\m}(p_1)&=&\bar{u}(p_1,s_1) \g^\m \left(\frac{\slashed{p_1}-\slashed{k}+M}{(p_1-k)^2-M^2+i\e}\right)\label{eq:fermion-line}\\F_1^{\m}(p_2)v(p_2,s_2)&=& \left(\frac{-\slashed{p_2}-\slashed{k}+M}{(p_2+k)-M^2}\right)\g^{\m}v(p_2,s_2)\nn\\
\bar{u}(p_1,s_1)\tilde{\Gamma}^{\rho}(p_1,p_2,M)v(p_2,s_2)&=&\int_k\bar{u}(p_1,s_1)F_1^{\m}(p_1)\g^{\r}F_1^{\n}(p_2)v(p_2,s_2)\frac{-ie^2\eta_{\mu\nu}}{k^2+i\e}. \nn
\eea
Here, the subscript indicates the number of gauge boson insertions on the fermion line.
In what follows, we will find it convenient to use the Grammer-Yennie decomposition \cite{Grammer:1973db} of the identity matrix,
\bea
\d^{\m}_{\;\n}&=&K^{\m}_{\;\n}(p_1,k)+G^{\m}_{\; \n}(p_1,k)\nn\\
K^{\m}_{\;\n}(p_1,k)&=&k_{\n}\frac{p_1^{\m}}{p_1\cdot k}\nn \\ 
G^{\m}_{\;\n}(p_1,k)&=&\d^{\m}_{\;\n}-k_{\n}\frac{p_1^{\m}}{p_1\cdot k}.\label{eq:G-K-decomp}
\eea 

Let us note that there is a freedom in the choice of the $K$ photon, with a numerator shifted by any vector proportional to $k^{\m}(O(1)+O(k)+\ldots)$ and the denominator shifted by $O(k^2)+\ldots$. Such shifts are higher order in $k$ and do not affect the leading power argument. However, at NLP, such a shift begins to contribute to the integrand, and we make a choice which separates the leading power from all non-leading contributions. This choice differs from that made in \cite{DelDuca:1990gz}. We now proceed to apply the K–G decomposition.
\bea
F_1^{\m}(p_1)&=&K^{\m}_{\; \nu}(p_1,k)F_1^{\n}(p_1)+G^{\m}_{\;\n}(p_1,k) F_1^{\n}(p_1),\nn \\ 
F_1^{\m}(p_2)&=&K^{\m}_{\; \nu}(p_2,k)F_1^{\n}(p_2)+G^{\m}_{\;\n}(p_2,k) F_1^{\n}(p_2). \label{eq:fermion-line-decomp}
\eea
Next, we observe that the G-photon satisfies the identity
\bea
G^{\m}_{\; \n}(p_1,k)p_1^{\n}=0,
\eea
which implies that the G-photon anti-commutes with $\slashed{p_1}$
\bea
G^{\mu}_{\;\n}\g^{\n}(\slashed{p_1}+M)=(-\slashed{p_1}+M)G^{\mu}_{\;\n}\g^{\n}.
\eea
As a result, the $\slashed{p_1}+M$ term in the first line of Eq.\ (\ref{eq:fermion-line}) and  the $\slashed{p_2}-M$ term in the second line identically vanish when $\g^{\m}$ is replaced by $G^{\m}_{\;\n}\g^{\n}$. We conclude that the G-photon is power suppressed relative to the K-photon. 

The K-photon is longitudinal and satisfies the Ward identity 
\bea
k_{\m}F_1^{\m}(p_1,k)&=&-1\nn \\ 
k_{\m}F_1^{\m}(p_2,k)&=&-1.\label{eq:ward-id}
\eea
As a result, the vertex correction with  $F_1^{\mu}(p_1,k)$ replaced by $K^{\m}_{\;\n}(p_1,k)F_1^{\n}(p_1,k)$, and $F_1^{\mu}(p_2,k)$ replaced by $K^{\m}_{\;\n}(p_2,k)F_1^{\n}(p_2,k)$ agrees with the leading power expression in Eq.\ (\ref{eq:one-loop-lp}).  Therefore, all next-to-leading power terms at one loop must come from the insertion of a single G-photon. The Ward identity used here involves no approximation. The leading power approximation is therefore neatly summarized by the assertion that all terms involving G-photons are to be dropped at leading power. 

We notice that  the vertex correction with the $F^{\mu}(p_1,k)$ replaced by $G^{\m}_{\;\n}(p_1,k)F_1^{\n}(p_1,k)$, and $F_1^{\mu}(p_2,k)$ replaced by $G^{\m}_{\;\n}(p_2,k)F_1^{\n}(p_2,k)$ is next-to-next-to leading power and we may absorb it into the hard part when working at next-to-leading power. We may now write
\bea
\tilde{\Gamma}^{\rho,\, (1)}_{\text{NLP}}(p_1,p_2,M)&=& \int_k \left[G^{\m}_{\;\a}(p_1,k)F_1^{\a}(p_1,k)\g^{\r}K^{\n}_{\;\b}(p_2,k)F_1^{\b}(p_2,k)\right. \nn \\
&&\left.+K^{\m}_{\;\a}(p_1,k)F_1^{\a}(p_1,k)\g^{\r}G^{\n}_{\;\b}(p_2,k)F_1^{\b}(p_2,k)\right]\frac{-ie^2\eta_{\mu\nu}}{k^2+i\e}.\label{eq:NLP-G-K-form}
\eea
Let us briefly comment on the qualitative features of G-K decomposition. The sum of the $G,K$ photons is the identity, and the $K$ is to be interpreted as the photon connected to an eikonal (Wilson) line. Therefore, the G photon is to be identified as being the remainder after subtraction of the eikonal piece. It contains sub-eikonal corrections as well as numerator corrections to the eikonal approximation. At leading power, all G photon attachments are absorbed into the hard function. At NLP, only one G-photon escapes from the hard part (at one loop), while the rest remain within the definition of the hard function. 

Finally, we would like to write an explicit expression for $G^{\m}_{\;\a}(p_1,k)F_1^{\a}(p_1,k)$
\bea
G^{\m}_{\;\a}(p_1,k)F_1^{\a}(p_1,k)&=& \left(\g^{\m}-\frac{\slashed{k}p_1^{\m}}{p_1\cdot k}\right)\frac{-\slashed{k}}{(p_1-k)^2-M^2+i\e}\nn \\ &\approx& \left(\g^{\m}-\frac{\slashed{k}2p_1^{\m}}{2p_1\cdot k}\right)\frac{\slashed{k}}{2p_1\cdot k-i\e},\label{eq:G-photon-approx}
\eea
where the approximation in the second line retains the leading term in the denominator, since a non-leading term has already appeared in the numerator.  We can put the G-photon in a more familiar form using 
\bea
\left(\g^{\m}-\frac{\slashed{k}2p_1^{\m}}{2p_1\cdot k}\right)\frac{\slashed{k}}{2p_1\cdot k}=\frac{(k_\a\d^{\m}_\b-k_{\b}\d^{\m}_{\a})}{2p_1\cdot k}\g^{\b}\slashed{k}\frac{2p_1^{\a}}{2p_1\cdot k}.\label{eq:F_mn-identity}
\eea
The term inside the brackets on the right-hand side of the Eq.\ (\ref{eq:F_mn-identity}) , we may identify as the momentum space representation of the field strength tensor $\tilde{F}_{\m\n}(k)$. 

As a result, we may put one of the NLP terms into the matrix element form
\bea
\int_k G^{\m}_{\;\a}(p_1,k)F_1^{\a}(p_1,k)\g^{\r}K^{\n}_{\;\b}(p_2,k)F_1^{\b}(p_2,k)&=&\nn \\ &&\hspace{-7cm}\int_{-\infty}^{0}d\s_1\int_{-\infty}^{\s_1}d\s\frac{ev_1^2v^\a_1}{2p_1\cdot v_1}\frac{\left\langle \partial_{\n}F_{\a\m}(v_1\s) W_{v_1}(0,\infty)W_{v_2}(0,\infty)\right\rangle}{\langle W_{v_1}(0,\infty)W_{v_2}(0,\infty)\rangle} \g^{\m}\g^{\n}\g^{\r},\label{eq:Matrix-element-1}
\eea
where we have replaced the K-photon on the $p_2$ line by a Wilson line using the Ward identity in Eq.\ (\ref{eq:ward-id}) as in Eq.\ (\ref{eq:one-loop-lp}).  The single G-photon corresponds to the insertion of a single $F$ in the matrix element. The Wilson lines capture all K-photons insertions, following the leading power result. 

The insertion of the Wilson line $W_{v_1}(0,\infty)$ adds  the leading power graph, and the denominator subtracts the connected, leading power ladder graph from the matrix element. This has been done so as to ensure that the NLP soft function generalizes to higher orders with more complicated connected graphs. In particular, disconnected graphs cancel between the numerator and denominator. The numerator in Eq.\ (\ref{eq:Matrix-element-1})  is the position space analogue of Eq.\ (\ref{eq:F_mn-identity}). 

A similar analysis applies to the second term in Eq.\ (\ref{eq:NLP-G-K-form}), and we write all next-to-leading power terms at one loop in the factorized matrix element form
 \bea
 \tilde{\Gamma}^{\rho,\, (1)}_{\text{NLP}}(p_1,p_2,M)&=&\int_{-\infty}^{0}d\s_1\int_{-\infty}^{\s_1}d\s\frac{ev_1^2v^\a_1}{2p_1\cdot v_1}\frac{\left\langle \partial_{\n}F_{\a\m}(v_1\s) W_{v_1}(0,\infty)W_{v_2}(0,\infty)\right\rangle}{\langle W_{v_1}(0,\infty)W_{v_2}(0,\infty)\rangle} \g^{\m}\g^{\n}\G^{\r,(0)}\nn\\
 &&\hspace{-2cm}+\int^{\infty}_{0}d\s_1\int_{0}^{\s_1}d\s\frac{-ev_2^2v^\a_2}{2p_2\cdot v_2}\frac{\left\langle \partial_{\n}F_{\a\m}(v_2\s) W_{v_1}(0,\infty)W_{v_2}(0,\infty)\right\rangle}{\langle W_{v_1}(0,\infty)W_{v_2}(0,\infty)\rangle} \G^{\r,(0)}\g^{\n}\g^{\m}.\label{eq:one-loop-NLP}
 \eea
 Therefore, we have shown that at one-loop, the factorization is preserved at NLP and that the long-distance function involves an insertion of a gauge-invariant field strength. In summary, the new one-loop factorization reads,
\bea
\tilde{\Gamma}^{\r,(1)}=\tilde{\Gamma}^{\r,(0)}\langle W_{v_1}(0,\infty)W_{v_2}(0,\infty)\rangle +\tilde{\Gamma}^{\r,(1)}_{\text{NLP}}+\tilde{H}^{\r,(1)}(p_1,p_2),\label{eq:one-loop-factorization}\label{eq:one-loop-result}
\eea  
where we have $\tilde{\Gamma}^{\r,(1)}_{\text{NLP}}$ is the function that appears in the right hand side of Eq.\ (\ref{eq:one-loop-NLP}), and remainder $\tilde{H}$ defines a new hard function where both leading power and next-to-leading power long distance physics has been factored out. The full form factor has a vertex correction counter term insertion which involves no loop integrals, and therefore trivially shifts the hard function that appears on the right-hand side of Eq.\ (\ref{eq:one-loop-result}) in a scheme-dependent way. The self-energy graphs that we have not studied here can be chosen to vanish on the mass shell by renormalizing on-shell \cite{Sterman:1993hfp}.

In the next section, we show that the one-loop factorization formula in Eq.\ (\ref{eq:one-loop-factorization}) does not hold at two loops. However, the two-loop NLP corrections factorize as a differential operator. 
 \section{Two-loop analysis}
 \label{sec:two-loop-analysis}
 In this section, we study next-to-leading power (NLP) factorization at two loops. As previously indicated, new terms arise at two loops that do not appear in the one-loop analysis. These are of two types: (i) a soft photon emerging from the hard part, which factors as a derivative operator acting on the lower-order hard function, and (ii) a double G-photon insertion on a fermion line. These terms are absent at one loop because the tree-level hard function is trivial, so its derivative vanishes. Excluding self-energy corrections on the external lines, at most one gauge boson attaches to an outgoing fermion, so multi-G-photon terms do not contribute at one loop.
 
 As before, instead of studying all graphs, we start by studying a representative subset. We will include other graphs that contribute to the form factor subsequently.
 We start our two-loop analysis by focusing on ladder-like graphs, denoted by $\tilde{\Gamma}^{\rho,(2)}$. The full form factor ${\Gamma}^{\rho,(2)}$ also includes additional diagrams, which we omit in this discussion.  
 
  \begin{figure}[h]
 	\begin{center}
 		\begin{tabular}{cc}
 			\includegraphics[width=.3\textwidth]{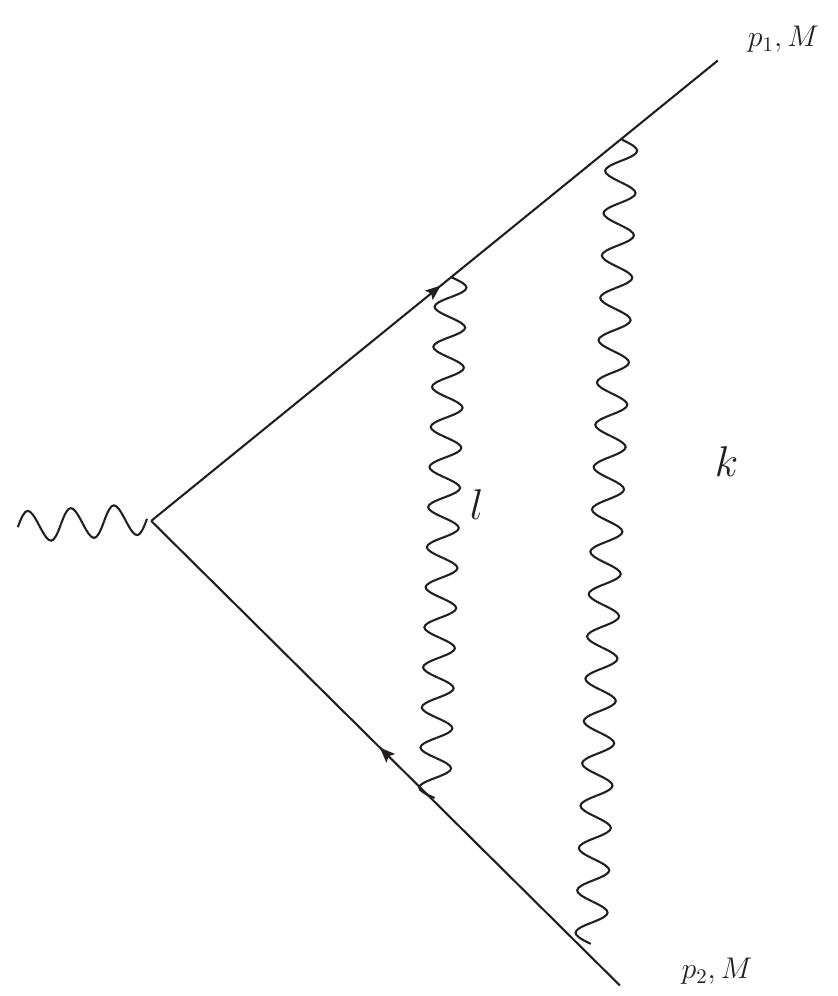}\hfill &
 			\includegraphics[width=.3\textwidth]{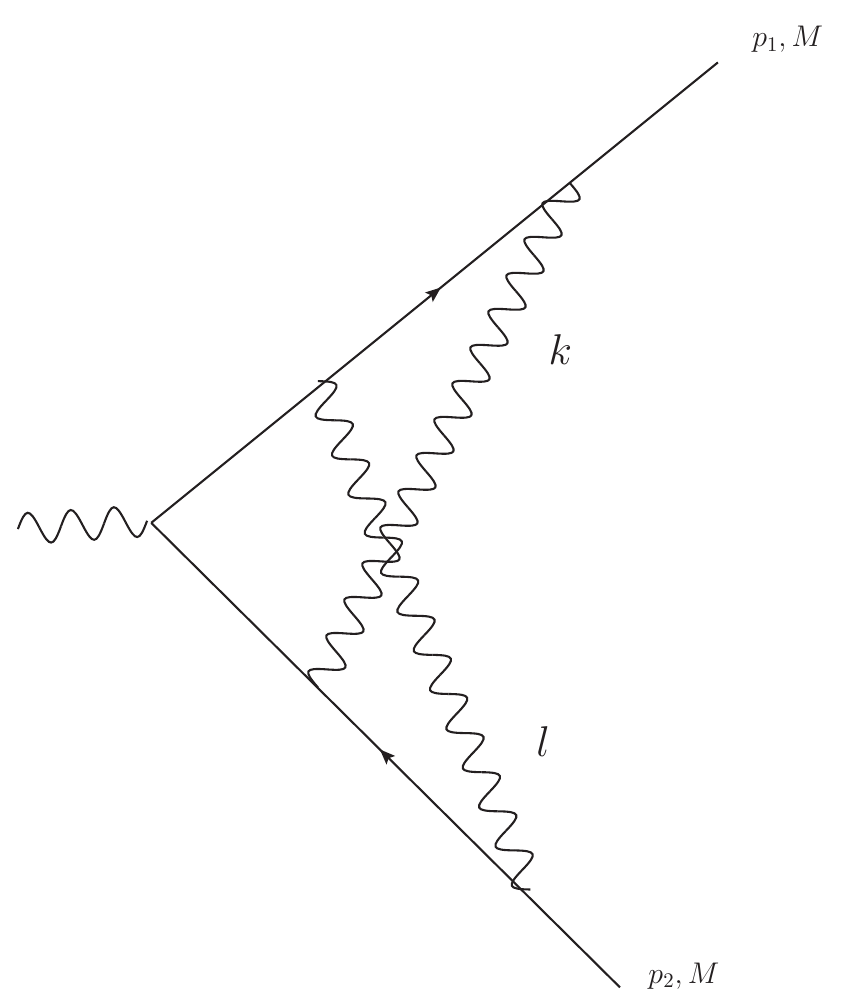}\hfill \\ 
 			(a) &(b) 
 		\end{tabular}
 		\caption{ Graphs that contribute to the two-loop form factor.  (a) Ladder graph (b) Crossed ladder graph } 
 		\label{fig:twoloop-graphs}
 	\end{center}
 \end{figure}
 
 The relevant graphs at two loops are shown in Fig.\ \ref{fig:twoloop-graphs}. Additional contributions include self-energy diagrams, nontrivial counterterm insertions, and vertex correction ladders. Closed fermion loops at this order contain three-photon subgraphs, which vanish by Furry’s theorem. In the presence of other light fermions, the self-energy with a fermion loop is also pinched. 
 
 Let us list all pinch surfaces of the two graphs in question. 
 \begin{itemize}
 	\item Both $l,k$ are soft: this region contributes a new term involving a double G-photon insertion.
 	\item The loop momentum $l$ is soft and  $k$ hard: this region is doubly power suppressed in the ladder diagram but contributes at NLP in the crossed-ladder case. We will obtain a $p_1$  derivative operator acting on the one-loop hard function. 
 	\item The loop momentum $l$ is hard and  $k$ soft: this region contributes at both leading and next-to-leading power in the ladder diagram. These contributions are encapsulated by Eq.\ (\ref{eq:one-loop-result}), albeit with a more intricate hard function. In the crossed-ladder case, this configuration again yields a $p_2$ derivative acting on the one-loop hard function.
 \end{itemize}
 
We begin by examining the double-soft limit. As in the one-loop case, we decompose both the ladder and crossed-ladder diagrams into contributions from two fermion lines.
\bea
\bar{u}(p_1,s_1)F_2^{\m_1\m_2}(p_1)&=&\bar{u}(p_1,s_1) \left(\g^{\m_1}\frac{\slashed{p}_1-\slashed{k}+M}{(p_1-k)^2-M^2+i\e}\g^{\m_2}\frac{\slashed{p}_1-\slashed{k}-\slashed{l}+M}{(p_1-l-k)^2-M^2+i\e}\right. \nn \\ &&+ \left. \g^{\m_2}\frac{\slashed{p}_1-\slashed{l}+M}{(p_1-l)^2-M^2+i\e}\g^{\m_1}\frac{\slashed{p}_1-\slashed{k}-\slashed{l}+M}{(p_1-l-k)^2-M^2+i\e} \right),
\eea
 and
 \bea
 F_2^{\n_1\n_2}(p_2) v(p_2,s_2)&=& \left(\frac{-(\slashed{p}_2+\slashed{k}+\slashed{l})+M}{(p_2+l+k)^2-M^2+i\e}\g^{\n_2}\frac{-(\slashed{p}_2+\slashed{k})+M}{(p_2+k)^2-M^2+i\e}\g^{\n_1}\right. \nn \\ && \hspace{-2cm}+ \left.  \frac{-(\slashed{p}_2+\slashed{k}+\slashed{l})+M}{(p_2+l+k)^2-M^2+i\e}\g^{\n_1}\frac{-(\slashed{p}_2+\slashed{l})+M}{(p_2+l)^2-M^2+i\e}\g^{\n_2} \right)v(p_2,s_2).
 \eea
 From these expressions, we can reconstruct the two ladder-like graphs using
 \bea
 \tilde{\Gamma}^{\r,(1)}(p_1,p_2)=\frac{e^4}{2}\int_{l,k} F_2^{\m_1\m_2}(p_1)\g^{\r}F_2^{\n_1\n_2}(p_2)  \frac{-i\eta_{\m_1\n_1}}{k^2+i\e}
\frac{-i\eta_{\m_2\n_2}}{l^2+i\e}.\label{eq:ladder-like}
 \eea
 Here, the factor of $\frac{1}{2}$ accounts for symmetrization over the loop momenta $k,l$.
 We can now carry out the  K-G decomposition on each fermion line independently as in Eq.\ (\ref{eq:fermion-line-decomp}). 
 \bea
 F_2^{\m_1\m_2}(p_1)&=&\left(K^{\m_1}_{\;\a_1}(p_1,k)K^{\m_2}_{\;\a_2}(p_1,l)+K^{\m_1}_{\;\a_1}(p_1,k)G^{\m_2}_{\;\a_2}(p_1,l)\right)F_2^{\a_1\a_2}(p_1)\nn\\ &&+\left(G^{\m_1}_{\;\a_1}(p_1,k)K^{\m_2}_{\;\a_2}(p_1,l)+G^{\m_1}_{\;\a_1}(p_1,k)G^{\m_2}_{\;\a_2}(p_1,l)\right)F_2^{\a_1\a_2}(p_1)\label{eq:2loop-decomp}
 \eea 
 It is easy to check that each K-photon satisfies a Ward identity similar to Eq.\ (\ref{eq:ward-id}).
 \bea
 K^{\m_1}_{\;\a_1}(p_1,k)F_2^{\a_1\a_2}=F_1^{\a_2}(p_1,l)\frac{-p_1^{\m_1}}{p_1\cdot k},\nn \\ 
 K^{\m_2}_{\;\a_2}(p_1,l)F_2^{\a_1\a_2}=F_1^{\a_1}(p_1,k)\frac{-p_1^{\m_2}}{p_1\cdot l}.\label{eq:ward-id-2}
 \eea 
 The structure of this Ward identity is $K\otimes F_n(p_i) =F_{n-1}(p_i)$, where $n$ is the number of attachments on the fermion line $p_i$. Therefore, the Ward identity makes it possible to peel off each longitudinal photon one at a time, in a recursive fashion. 
 
 We now examine each term in  Eq.\ (\ref{eq:2loop-decomp}) one at a time
 \bea
(K^{\m_1}_{\;\a_1}(p_1,k)K^{\m_2}_{\;\a_2}(p_1,l))F_2^{\a_1\a_2} &=& \frac{-p_1^{\m_1}}{p_1\cdot k}\frac{-p_1^{\m_2}}{p_1\cdot l},
 \eea
 which corresponds to the $O(e^2)$ expansion of the $p_1$ Wilson line in Eq.\ (\ref{eq:heavy-quark-LP}). 
 Next, we consider the $KG$ terms in Eq.\ (\ref{eq:2loop-decomp}). We apply the Ward identity in Eq.\ (\ref{eq:ward-id-2}) to obtain
 \bea
 (K^{\m_1}_{\;\a_1}(p_1,k)G^{\m_2}_{\;\a_2}(p_1,l))F_2^{\a_1\a_2}&=& \frac{-p_1^{\m_1}}{p_1\cdot k}G^{\m_2}_{\;\a_2}(p_1,l)F_1^{\a_2}(p_1,l),\nn \\ 
 (G^{\m_1}_{\;\a_1}(p_1,k)K^{\m_2}_{\;\a_2}(p_1,l))F_2^{\a_1\a_2}&=& \frac{-p_1^{\m_2}}{p_1\cdot l}G^{\m_1}_{\;\a_1}(p_1,k)F_1^{\a_1}(p_1,k).
 \eea
 The action of G-photons on $F_1$ has already been analyzed in Eqs.\ (\ref{eq:G-photon-approx}) and (\ref{eq:F_mn-identity}). The factor of $\frac{1}{2}$ in Eq.\ (\ref{eq:ladder-like}) cancels because the $k,l$ symmetry gives two equivalent ways to assign G-polarization to the photons on the $p_1$ line. Therefore, this corresponds to the expansion of the $p_1$ Wilson line to $O(e)$ and the momentum space analogue of $\partial F_{\m\n}$ operator in Eq.\ (\ref{eq:Matrix-element-1}). 
 
As a result, any genuinely new contribution to the NLP factorization theorem must arise from the double G-photon term, which was absent at one loop. Let us analyze these terms carefully. 
 \bea
 \bar{u}(p_1,s_1)G^{\m_1}_{\;\a_1}(p_1,k)G^{\m_2}_{\;\a_2}(p_1,l)F_2^{\a_1\a_2}(p_1)&=&G^{\m_1}_{\;\a_1}(p_1,k)G^{\m_2}_{\;\a_2}(p_1,l)\nn \\ &&\hspace{-4cm}\bar{u}(p_1,s_1) \left(\g^{\a_1}\frac{\slashed{p}_1-\slashed{k}+M}{(p_1-k)^2-M^2+i\e}\g^{\a_2}\frac{\slashed{p}_1-\slashed{k}-\slashed{l}+M}{(p_1-l-k)^2-M^2+i\e}\right. \nn \\ &&
 \hspace{-4cm}+ \left. \g^{\a_2}\frac{\slashed{p}_1-\slashed{l}+M}{(p_1-l)^2-M^2+i\e}\g^{\a_1}\frac{\slashed{p}_1-\slashed{k}-\slashed{l}+M}{(p_1-l-k)^2-M^2+i\e} \right).
 \eea
 In picking out the relevant pieces at NLP, we first observe that in either term, choosing $\slashed{p}_1+M$ in the first numerator (from the left) yields zero identically since $\slashed{p}_1$ anticommutes with $G^{\m_i}_{\a_i}\g^{\a_i}$, and the Dirac equation implies that this term vanishes when acting on the spinor. Further, choosing $-\slashed{k}-\slashed{l}$ in the second numerator (from the left)next-to-next-to-leading power ($\text{N}^2$LP). For the same reason, we apply the eikonal approximation to both denominators.
  After making the relevant approximations, we obtain  
 \bea
 G^{\m_1}_{\;\a_1}(p_1,k)G^{\m_2}_{\;\a_2}(p_1,l)F_2^{\a_1\a_2}(p_1)&\approx&G^{\m_1}_{\;\a_1}(p_1,k)G^{\m_2}_{\;\a_2}(p_1,l)\nn \\ &&\hspace{-4cm} \left(\g^{\a_1}\frac{-\slashed{k}}{-2p_1\cdot k+i\e}\g^{\a_2}\frac{\slashed{p}_1+M}{-2p_1\cdot(l+k)+i\e}\right. \nn \\ &&
 \hspace{-4cm}+ \left. \g^{\a_2}\frac{-\slashed{l}}{-2p_1\cdot l+i\e}\g^{\a_1}\frac{\slashed{p}_1+M}{-2p_1\cdot (l+k)+i\e} \right).
 \eea
 We may anticommute the rightmost $\slashed{p_1}+M$ past the G-polarized $\g$ matrices to give zero by use of the Dirac equation once again. The only nonvanishing term arises from the inner product of $p_1$ with the soft momentum. 
 \bea
 G^{\m_1}_{\;\a_1}(p_1,k)G^{\m_2}_{\;\a_2}(p_1,l)F_2^{\a_1\a_2}(p_1)&\approx&G^{\m_1}_{\;\a_1}(p_1,k)G^{\m_2}_{\;\a_2}(p_1,l)
\frac{\eta^{\a_1\a_2}}{2p_1\cdot (l+k)},
 \eea
 where the Dirac matrices reduce to the metric tensor $\eta^{\a_1\a_2}$. The G-photons satisfy an identity that manifests a gauge invariant coupling between G-photons and the fermion line
 \bea
  \e_{\m}(k)G^{\m}_{\;\a}(p_1,k)&=& \tilde{F}_{\b\a}(k)\frac{p^{\b}_1}{p_1\cdot k},\nn \\ 
  \tilde{F}_{\b\a}(k)&=&k_\b\e_\a- k_\b\e_\a.\label{eq:field-strength-G}
 \eea 
 As a result, we can write the new term as
 \bea
 G^{\m_1}_{\;\a_1}(p_1,k)G^{\m_2}_{\;\a_2}(p_1,l)
 \frac{\eta^{\a_1\a_2}}{2p_1\cdot (l+k)}&=&\frac{\eta^{\a_1\a_2}}{2p_1\cdot (l+k)} \frac{p_1^{\b_1}}{k\cdot p_1}
\frac{p_1^{\b_2}}{l\cdot p_1}\nn \\ &&\hspace{-2cm}\times\left(k_{\b_1}\d^{\m_1}_{\a_1}-k_{\a_1}\d^{\m_1}_{\b_1}\right)
\left(l_{\b_2}\d^{\m_2}_{\a_2}-l_{\a_2}\d^{\m_2}_{\b_2}\right).
 \eea
 Finally, we write a position space matrix element as in Eq.\ (\ref{eq:Matrix-element-1}) for the double G-photon insertion. 
 \bea
  \frac{1}{2}\int_{l,k}G^{\m_1}_{\;\a_1}(p_1,k)
  G^{\m_2}_{\;\a_2}(p_1,l)
  F_2^{\a_1\a_2}(p_1)\g^{\rho}
  K^{\n_1}_{\;\b_1}(p_2,k)K^{\n_2}_{\;\b_2}(p_2,l)F_2^{\b_1\b_2}(p_2) =\g^{\rho}\nn \\ \times \int\limits_{-\infty}^0 d\s_0 \int\limits_{-\infty}^{\s_0} d\s_1  \int\limits_{-\infty}^{\s_0} d\s_2 \frac{e^2 v_1^2 v_1^{\a}v_1^{\b}\eta^{\m\n}}{4p_1\cdot v_1} \frac{\left\langle F_{\a\m}(v_1\s_1)F_{\b\n}(v_1\s_2) W_{v_1}(0,\infty)W_{v_2}(0,\infty)\right\rangle}{\langle W_{v_1}(0,\infty)W_{v_2}(0,\infty)\rangle}
 \label{eq:Matrix-element-2}
 \eea
 Once again, the two factors of $F$ reproduce the double G photon insertion, while the Wilson lines capture all the K photons (of which there are infinite at all orders in perturbation theory).
 
\begin{figure}[h]
 	\begin{center}
 		\begin{tabular}{cc}
 			\includegraphics[width=.35\textwidth]{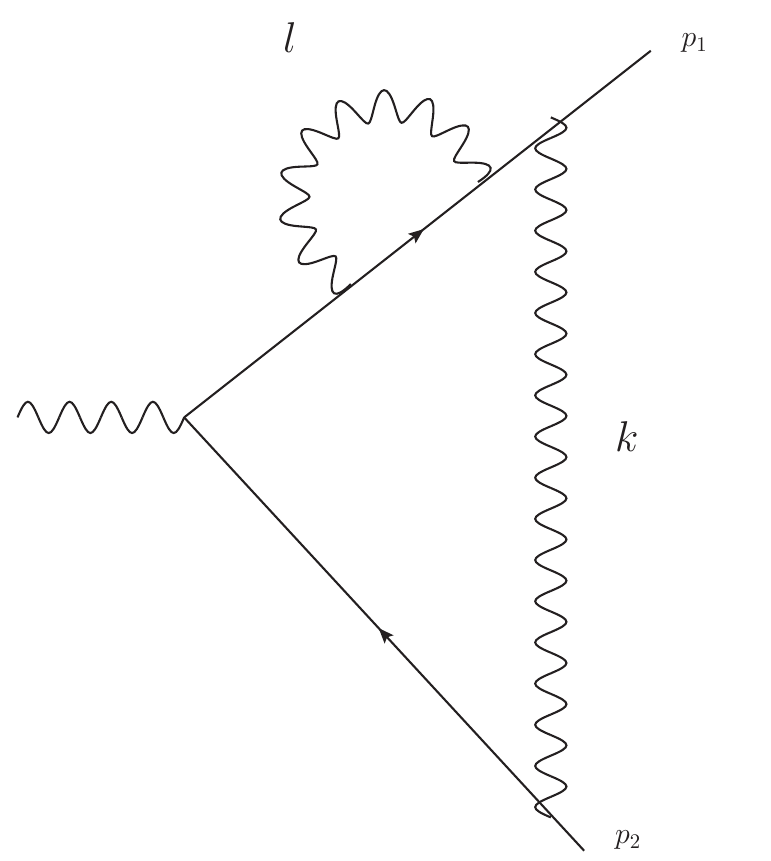}\hfill &
 			\includegraphics[width=.3\textwidth]{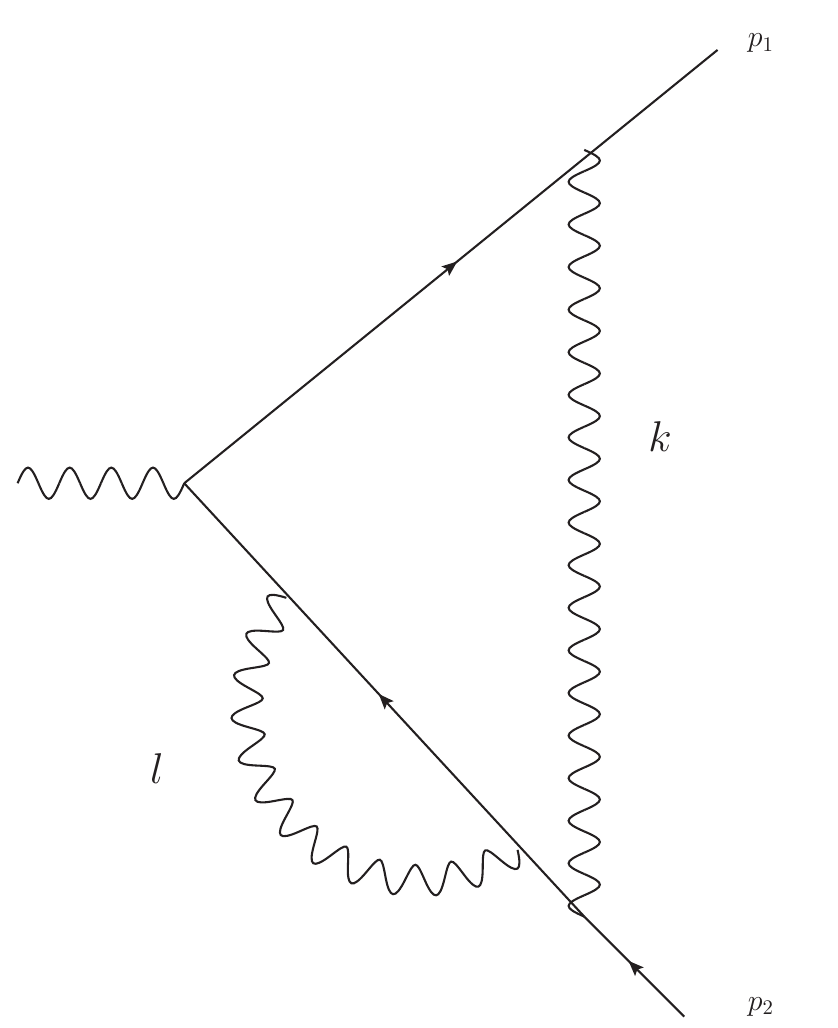}\hfill \\ 
 			(a) &(b) 
 		\end{tabular}
 		\caption{ Graphs that contribute to the two loop form factor but not at NLP when one loop is soft and the other is hard.  (a) The $p_1$ self-energy ladder graph  (b) The $p_2$ self-energy ladder graph  } 
 		\label{fig:self-energy-ladder}
 	\end{center}
 \end{figure}

 Let us now turn to the remaining pinch surfaces. To derive the NLP factorization theorem for the form factor in the other regions, we must include additional diagrams that contribute at this order.
  First, consider the self energy ladders in Fig.\ \ref{fig:self-energy-ladder}. 
 
 Consider the graph with $p_1$ self-energy ladder shown in Fig.\ \ref{fig:self-energy-ladder}(a). When $k$ is soft and $l$ is hard, the leading-power contribution from the soft photon attachment to the fermion lines can be replaced by their corresponding K-photon components.
  When combined with other attachments on the $p_1$ line, a self-energy diagram remains, which vanishes under on-shell renormalization. When the soft photon, labelled by the loop momentum $k$, is G-polarized, the $k$ integral factorizes from the self-energy,  yielding a matrix element in the form of Eq.\ (\ref{eq:Matrix-element-1}). This leaves behind an on-shell self-energy diagram, which again vanishes upon renormalization.

 When $l$ is soft and $k$ is hard, both self-energy ladders are doubly power suppressed because they contain two fewer eikonal lines.
 
\begin{figure}[h]
 	\begin{center}
 		\begin{tabular}{cc}
 			\includegraphics[width=.35\textwidth]{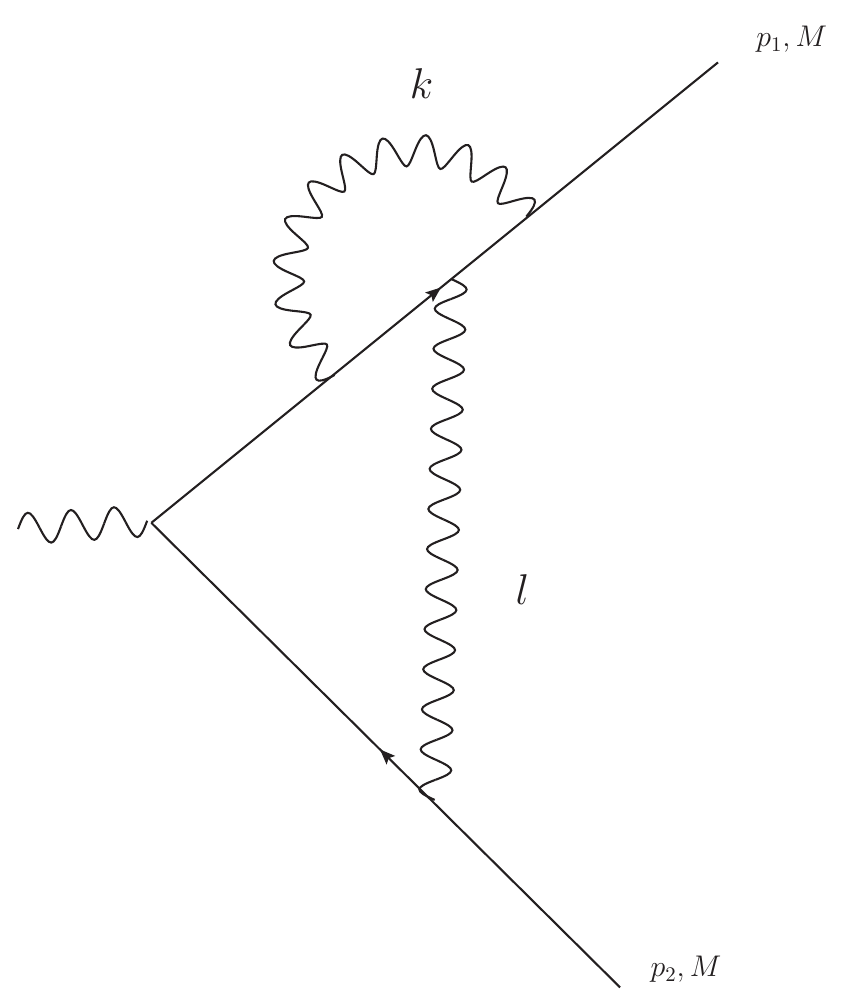}\hfill &
 			\includegraphics[width=.3\textwidth]{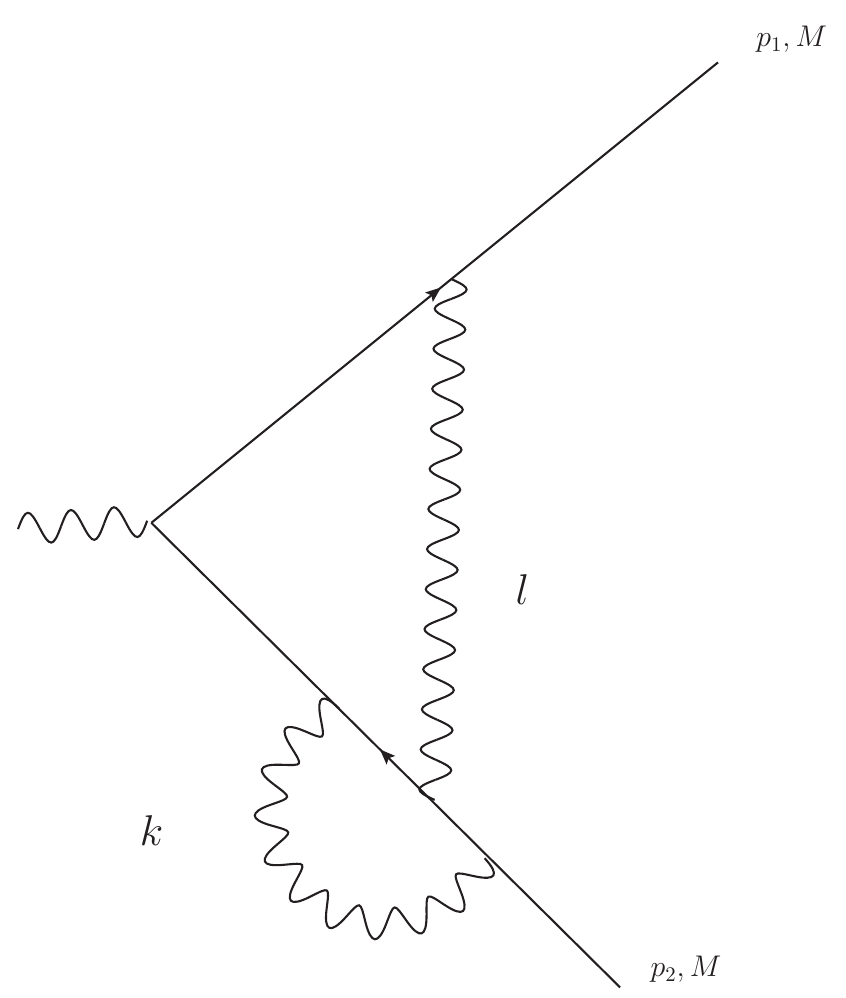}\hfill \\ 
 			(a) &(b) 
 		\end{tabular}
 		\caption{ Graphs that contribute to the two-loop form factor, at NLP through the vertex correction to the ladder exchange as well as through soft momentum entering the hard part  (a) The $p_1$ vertex correction ladder graph  (b) The $p_2$ vertex correction ladder graph  } 
 		\label{fig:vertex-correction-ladder}
 	\end{center}
 \end{figure}

 When $l$ and $k$ are both soft, the G photon contributions are non-trivial, but as we will see in Sec.\ \ref{sec:all-order-analysis}, such contributions are all captured by Eqs.\ (\ref{eq:Matrix-element-1}) and (\ref{eq:Matrix-element-2}). 
 Next, we consider the graphs in Fig.\ \ref{fig:vertex-correction-ladder}.
 
 The vertex-correction ladder diagrams possess a double-soft region, where both $k$ and $l$ are soft. In this region, at leading power, we may replace each attachment to the fermion lines by their K-photon components. At next-to-leading power, one or two attachments can be G photons instead, and the relevant graphs are generated $O(e^4)$ expansion of Eqs.\ (\ref{eq:Matrix-element-1}) and (\ref{eq:Matrix-element-2}). As we will argue in Sec.\ \ref{sec:all-order-analysis}, the case of three or more G-polarized photons is a next-to-next-to-leading power (N$^2$LP) effect. 
 
  \begin{figure}[h]
 	\begin{center}
 		\begin{tabular}{cc}
 			\includegraphics[width=.30\textwidth]{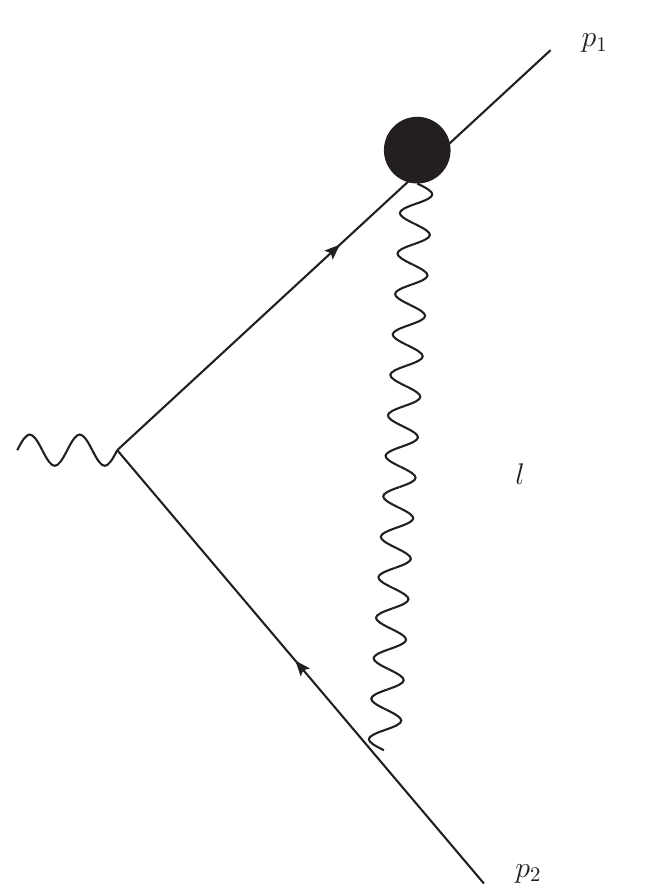}\hfill &
 			\includegraphics[width=.36\textwidth]{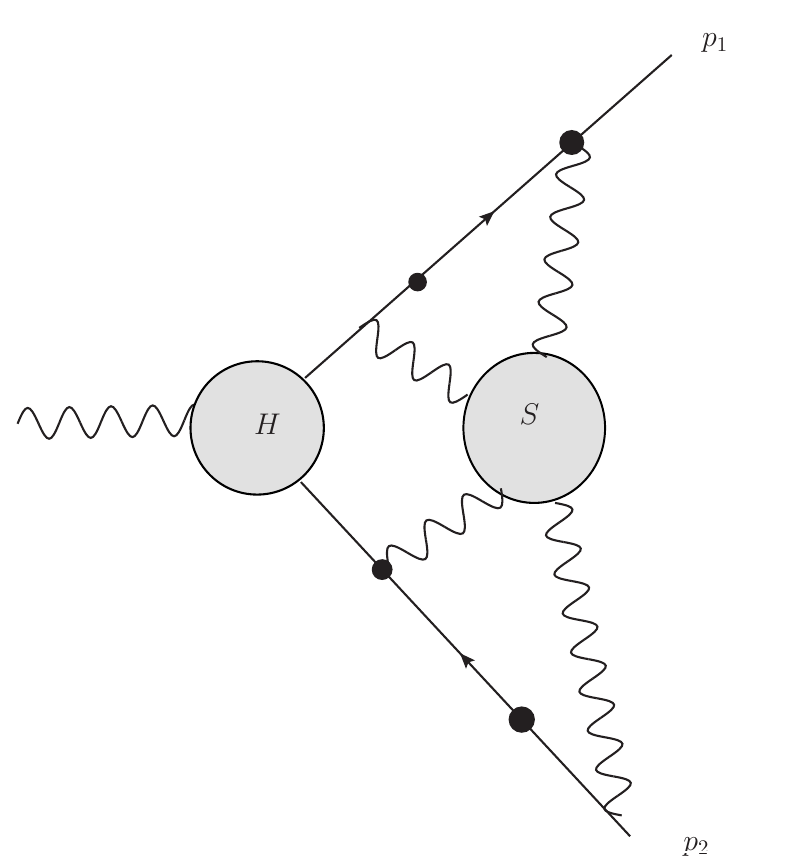}\hfill \\ 
 			(a) &(b) 
 		\end{tabular}
 		\caption{ Pinch surfaces with composite vertex insertions are non-trivial at next-to-leading power (a) The two-loop composite vertex ladder when $l$ is soft. The vertex correction in Fig.\ \ref{fig:vertex-correction-ladder}  has been replaced by a black dot.   (b) A general reduced graph with composite vertex insertions } 
 		\label{fig:composite-vertex-ladder}
 	\end{center}
 \end{figure}
 Next, we study the possibility that $l$ is soft. In this situation, we arrive at a new pinch surface, not displayed in Fig.\ \ref{fig:pinches}. These are pinch surfaces with composite hard vertices. The situation at two loops is  shown in Fig.\ \ref{fig:composite-vertex-ladder} (a), while a more general pinch surface involving composite hard vertices appears in Fig.\ \ref{fig:composite-vertex-ladder} (b). The relevant composite vertices are two-point and three-point interactions containing internal hard subgraphs. At leading power, all such composite vertices reduce to their tree-level values, as the hard subgraphs can be approximated by their zero-momentum limits. In an on-shell scheme, the on-shell self-energy vanishes, and the zero-momentum limit of the three-point function coincides with its tree-level value by definition.
 Therefore, all such composite vertices may be replaced by tree vertices for leading power calculations. However, at next-to-leading power, the vertex must be expanded to first order in the soft momentum. For particles with spin one-half, such an expansion coincides with the perturbative expansion for the anomalous magnetic moment. Therefore, we may write
 \bea
 \Gamma^\r_{\text{vc-ladder,l soft,NLP}}&=&\int\limits_{l\in \text{Soft}} \frac{-ie\eta_{\m\n}}{l^2+i\e} \frac{-p_2^{\n}}{p_2\cdot l+i\e} ie\sigma^{\mu\a}\frac{l_{\a}}{M}F_2(0) \frac{i(\slashed{p}_1+M)}{-2p_1\cdot l+i\e}\g^{\r}\nn \\ && +\frac{-ie\eta_{\m\n}}{l^2+i\e} \frac{p_1^{\m}}{-p_1\cdot l+i\e}\g^{\r}\frac{i(-\slashed{p}_2+M)}{2p_2\cdot l+i\e}ie\sigma^{\nu\a}\frac{-l_{\a}}{M}F_2(0)\label{eq:composite-hard}
 \eea
 Here, the Dirac matrices $\s^{\m\n}=\frac{i}{2}\left[\g^{\m},\g^{\n}\right]$ are the generators of Lorentz boosts, formed from the commutator of gamma matrices. The scalar factor $F_2(0)$ is the zero momentum form factor, which corresponds to the anomalous magnetic moment of the electron. This function is ultraviolet (UV) finite to all orders in perturbation theory, since QED permits no counterterm proportional to $\s^{\m\n}$.  It is also infrared (IR) finite to all orders. Therefore, $F_2(0)$ admits a well-defined expansion in $\a$. 
 
 In particular, the composite hard vertex takes the explicit form
 \bea
 F_2(0)=\frac{\a}{24\pi}+O(\a^2).
 \eea
The expansion for the zero-momentum form factor coincides with the electron’s anomalous magnetic moment. The final observation we make about the composite hard vertex in Eq.\ (\ref{eq:composite-hard}) is that it does not couple to $K$ photons as a result of the identity $\s^{\m\n}l_\m l_\n=0$. It couples to the field strength through the identity in Eq.\ (\ref{eq:field-strength-G}). We may therefore write Eq.\ (\ref{eq:composite-hard}) in a matrix element form:
 \bea
 \Gamma^{\r,(2)}_{\text{vc-ladder,l soft,NLP}}&=&F_2^{(1)}(0) \left[\int_{-\infty}^{0}d\s_1\int_{-\infty}^{\s_1}d\s \frac{iev_1^2v_1^{\a}}{M p_1\cdot v_1} \right.  \\ && \left. \times \frac{\langle \s^{\m\n}\partial_{\n}F_{\a\m}(\s v_1)W_{v_1}(0,\infty) W_{v_2}(0,\infty)\rangle}{\langle W_{v_1}(0,\infty)W_{v_2}(0,\infty)\rangle}i(\slashed{p}_1+M)\G^{\r,(0)}\right. \nn \\ 
 &&\hspace{-4cm}\left.+\G^{\r,(0)} i(\slashed{p}_2+M)\int^{\infty}_{0}d\s_1\int_{0}^{\s_1}d\s \frac{-iev_2^2v_2^{\a}}{M p_2\cdot v_2}
\frac{\langle \s^{\m\n}\partial_{\n}F_{\a\m}(\s v_2)W_{v_1}(0,\infty) W_{v_2}(0,\infty)\rangle}{\langle W_{v_1}(0,\infty)W_{v_2}(0,\infty)\rangle}
\right]\nn
 \eea 
 
 The last remaining region of the vertex correction ladders in Fig.\ \ref{fig:vertex-correction-ladder} is the $k$ soft, $l$ hard region. This configuration contributes at subleading power, and we now address it.
To analyze it, we must combine it with the ladder and crossed-ladder diagrams in  the $k$ soft, $l$ hard region. Qualitatively, this gives the fourth and final contribution to the NLP factorization theorem. In the ladder graph, only one soft parton attaches to the $p_1$ fermion line. We may therefore repeat the analysis of Sec.\ \ref{sec:one-loop-analysis} in the presence of a non-trivial hard function. In particular, the G photon terms in Eq.\ (\ref{eq:Matrix-element-1}) are unchanged. However, in the ladder diagram, the soft momentum also flows through the hard function. Therefore, the implicit leading-power assumption—that soft momenta can be neglected within the hard part—must be relaxed.
 
 We can begin our analysis by decomposing the ladder-like graphs into their individual contributions
 \bea
 \tilde{\Gamma}^{\rho}(p_1,p_2)={\Gamma}_{\text{lad}}^{\rho}(p_1,p_2)+{\Gamma}_{\text{c-lad}}^{\rho}(p_1,p_2).
 \eea
 
 In the region of interest, we may write
 \bea
 {\Gamma}_{\text{lad},\text{k-soft,l-hard}}^{\rho}(p_1,p_2)&=&\int\limits_{k\in \text{Soft}} F_1^{\mu}(p_1,k) H^{(1),\r}(p_1-k,p_2+k)F_1^{\nu}(p_2,k) \frac{-ie^2 \eta_{\m\n}}{k^2+i\e} \nn \\ 
 H^{(1),\r}(p_1-k,p_2+k)&=&\int\limits_{l\in \text{Hard}} \g^{\m_2}\frac{\slashed{p}_1-\slashed{k}-\slashed{l}+M}{(p_1-l-k)^2-M^2+i\e}\g^\r \label{eq:hard-function-ladder}
  \\ && \hspace{2cm}
 \times \frac{-(\slashed{p}_2+\slashed{k}+\slashed{l})+M}{(p_2+l+k)^2-M^2+i\e}\g^{\n_2} \frac{-ie^2 \eta_{\m_2\n_2}}{l^2+i\e}\nn 
 \eea
 At leading power, we replace $F_1$ by its  $K$ photon components and $ H^{\r}(p_1-k,p_2+k)$ by $ H^{\r}(p_1,p_2)$.   At NLP, we must consider the G-photon term as before, and repeating the analysis of Sec.\ \ref{sec:one-loop-analysis}, we arrive at the formula in Eq.\ (\ref{eq:Matrix-element-1}). However, we must also expand the hard function using
 \bea
  H^{(1),\r}(p_1-k,p_2+k)\approx H^{(1),\r}(p_1,p_2) +k^\a\sum_{i=1}^2 {-e_i}\frac{\partial H^{\r}(p_1,p_2)}{\partial p_i^\a}.
 \eea
 where we have defined $e_1=-e_2=1$. 
  To systematically consider the remaining regions, it is useful to study real radiation graphs at one loop and construct the two-loop form factor graphs from them using
 \bea
  \Gamma^{\r,(2)}_{\text{ladder,k soft,l hard,NLP}}+\Gamma^{\r,(2)}_{\text{S.E,k soft,l hard,NLP}}&=&\frac{1}{2}\int_{k_1\in \text{Soft}}\int_{k_2}(2\pi)^d\frac{-ie^2\eta_{\mu_1 \mu_2}\d^d(k_1+k_2)}{k_1^2+i\e}\nn \\
 && \hspace{-6cm}\times \,\sum_{j=1}^{2}\left( \sum_{i=1}^2e_ik_j^{\a}\frac{\partial H(p_1,p_2)}{\partial p_i^\a} \frac{p_i^{\m_j}}{p_i\cdot k_j+i\e}\right)\left(\sum_{i'=1}^2 \frac{e_{i'}p_{i'}^{\m_{\tilde{j}}}}{p_{i'}\cdot k_{\tilde{j}}+i\e}\right),\label{eq:2loop-sh-Kphoton}
 \eea  
where $\tilde{j}=2$ if $j=1$, and $\tilde{j}=1$ if $j=2$. We have included the self-energy graphs, which are zero in the on-shell scheme.

Yet another qualitatively new contribution comes from a soft photon entering the hard function. In two-loop graphs, these include the vertex correction ladder of Fig.\ \ref{fig:vertex-correction-ladder} in  the $k$ soft, $l$ hard region, as well as the crossed ladder in the $k$ soft, $l$ hard  and $k$ hard, $l$ soft regions.
\bea
 \Gamma^{\r,(2)}_{\text{c-ladder,k soft,l hard,NLP}}+ \Gamma^{\r,(2)}_{\text{c-ladder,l soft,k hard,NLP}}+ \Gamma^{\r,(2)}_{\text{vc-ladder,k soft,l hard,NLP}}=\nn \\
 \frac{1}{2}\int_{k_1\in \text{Soft}}\int_{k_2}(2\pi)^d\frac{-ie^2\eta_{\mu_1 \mu_2}\d^d(k_1+k_2)}{k_1^2+i\e}\left(\sum_{i'=1}^2 \frac{e_{i'}p_{i'}^{\m_{\tilde{j}}}}{p_{i'}\cdot k_{\tilde{j}}+i\e}\right)H_3^{(1),\r\m_j}(p_1,p_2,k_j), \label{eq:2loop-sh}
\eea
where we have introduced a new three particle hard function $H_3^{\r\n}(p_1,p_2,k)$, given by
\bea
H^{(1),\r\n}_3(p_1,p_2,k)&=&\int_{l\in\text{Hard}}\left(\g^\a\frac{\slashed{p_1}-\slashed{l}+M}{(p_1-l)^2-M^2+i\e}\g^\r\frac{-\slashed{p_2}-\slashed{l}-\slashed{k}+M}{(p_2+l+k)^2-M^2+i\e}\g^\n \right. \nn \\ 
&& \hspace{-2cm}\times\left.\frac{-\slashed{p_2}-\slashed{l}+M}{(p_2+l)^2-M^2+i\e}\g^{\b}\frac{-ie^2\eta_{\a\b}}{l^2+i\e}\right)+\left(\g^\a\frac{\slashed{p_1}-\slashed{l}+\slashed{k}+M}{(p_1-l+k)^2-M^2+i\e}\g^\n\right.\nn \\ 
&&\hspace{-2cm}\left. \times \frac{\slashed{p_1}-\slashed{l}+M}{(p_1-l)^2-M^2+i\e} \g^{\r}.\frac{-\slashed{p_2}-\slashed{l}+M}{(p_2+l)^2-M^2+i\e}\g^{\b}\frac{-ie^2\eta_{\a\b}}{l^2+i\e} \right).
\eea
At leading power, we apply Low's theorem \cite{Low:1958sn},\cite{DelDuca:1990gz} to write
\bea
H^{(1),\r\n}_3(p_1,p_2,k)=\sum_{i=1}^2-e_i\frac{\partial H^{(1),\r\n} (p_1,p_2)}{\partial p_i^{\n}}
\eea
Combining Eqs.\ (\ref{eq:2loop-sh-Kphoton}) and (\ref{eq:2loop-sh}) using Low's theorem, we may write
\bea
\Gamma^{\r,(2)}_{\text{c-ladder,k soft,l hard,NLP}}+ \Gamma^{\r,(2)}_{\text{c-ladder,l soft,k hard,NLP}}+ \Gamma^{\r,(2)}_{\text{vc-ladder,k soft,l hard,NLP}}\nn \\
+ \Gamma^{\r,(2)}_{\text{ladder,k soft,l hard,NLP}}+\Gamma^{\r,(2)}_{\text{S.E,NLP}}=\frac{1}{2}\int_{k_1\in \text{Soft}}\int_{k_2}(2\pi)^d\frac{-ie^2\eta_{\mu_1 \mu_2}\d^d(k_1+k_2)}{k_1^2+i\e}\nn \\
\times \,\sum_{j=1}^{2}\left( \sum_{i=1}^2 \left(k_{j\a}\d_{\b}^{\m_j}-k_{j\b}\d_{\a}^{\m_j}\right)\frac{\partial H(p_1,p_2)}{\partial p_i^\a} \frac{e_i p_i^{\b}}{p_i\cdot k_j+i\e}\right)\left(\sum_{i'=1}^2 \frac{e_{i'}p_{i'}^{\m_{\tilde{j}}}}{p_{i'}\cdot k_{\tilde{j}}+i\e}\right).
\eea

We observe that the $j=1,\tilde{j}=2$ term is exactly equal to the $j=2,\tilde{j}=1$ term by Bose symmetry. Therefore, we may select the $j=1,\tilde{j}=2$ term and write
\bea
\Gamma^{\r,(2)}_{\text{c-ladder,k soft,l hard,NLP}}+ \Gamma^{\r,(2)}_{\text{c-ladder,l soft,k hard,NLP}}+ \Gamma^{\r,(2)}_{\text{vc-ladder,k soft,l hard,NLP}}\nn \\
+ \Gamma^{\r,(2)}_{\text{ladder,k soft,l hard,NLP}}+\Gamma^{\r,(2)}_{\text{S.E,NLP}}=\int_{k\in \text{Soft}}\frac{-ie^2\eta_{\mu_1 \mu_2}}{k^2+i\e}\nn \\
\times\left( \sum_{i=1}^2 \left(k_{\a}\d_{\b}^{\m_1}-k_{\b}\d_{\a}^{\m_1}\right)\frac{\partial H(p_1,p_2)}{\partial p_i^\a} \frac{e_i p_i^{\b}}{p_i\cdot k+i\e}\right)\left(\sum_{i'=1}^2 \frac{e_{i'}p_{i'}^{\m_2}}{-p_{i'}\cdot k+i\e}\right).\label{eq:derivative-op}
\eea
We can now check that the self-energy term in this expression, which we added by hand, does indeed vanish.
\bea
\Gamma^{\r,(2)}_{\text{S.E,NLP}}&=&\int_{k\in \text{Soft}}\frac{-ie^2\eta_{\mu_1 \mu_2}}{k^2+i\e}k_{\a}\frac{\partial H(p_1,p_2)}{\partial p_1^{\a}} \left(\frac{ p_1^{\m_1}}{p_1\cdot k+i\e} \frac{p_{1}^{\m_2}}{-p_{1}\cdot k+i\e}\right)\nn\\
&=&0,
\eea
where the final equality follows from the fact that the integrand is odd under $k\rightarrow -k$.

Let us write a matrix element representation for the expression in Eq.\ (\ref{eq:derivative-op})
\bea
\Gamma^{\r,(2)}_{\text{c-ladder,k soft,l hard,NLP}}+ \Gamma^{\r,(2)}_{\text{c-ladder,l soft,k hard,NLP}}+ \Gamma^{\r,(2)}_{\text{vc-ladder,k soft,l hard,NLP}}&&\nn \\
+ \Gamma^{\r,(2)}_{\text{ladder,k soft,l hard,NLP}}+\Gamma^{\r,(2)}_{\text{S.E,NLP}}&&\nn \\ &&\hspace{-12cm}=\frac{\partial H^{(1),\rho}(p_1,p_2)}{\partial p_1^{\b}}\int_{-\infty}^{0}d\s \, {ev_1^{\a}}\frac{\langle F_{\b\a}(\s v_1)W_{v_1}(0,\infty) W_{v_2}(0,\infty)\rangle}{\langle W_{v_1}(0,\infty)W_{v_2}(0,\infty)\rangle} \nn \\ &&\hspace{-12cm} 
-\frac{\partial H^{(1),\rho}(p_1,p_2)}{\partial p_2^{\b}}\int^{\infty}_{0}d\s \, {ev_2^{\a}}\frac{\langle F_{\b\a}(\s v_2)W_{v_1}(0,\infty) W_{v_2}(0,\infty)\rangle}{\langle W_{v_1}(0,\infty)W_{v_2}(0,\infty)\rangle}
\eea
We may now collect all the NLP contributions at two loops and write
\bea
\Gamma^{\rho}&=&\langle W_{v_1}W_{v_2}\rangle\left(\tilde{H}^{\rho}+\sum_{i=1}^2\frac{\partial \tilde{H}^{\rho}(p_1,p_2)}{\partial p_i^{\b}}\int d\s \, {e_iev_i^{\a}}\frac{\langle F_{\b\a}(\s v_i) W_{v_2}W_{v_1}\rangle}{\langle W_{v_2}W_{v_1}\rangle}\right. \nn \\ 
&&\hspace{0cm}+\sum_{i=1}^{2}F_2(0) \left[\int d^2\s \frac{ie_iev_i^2v_i^{\a}}{M p_i\cdot v_i}\frac{\left\langle \left[\s^{\m\n},\left[i(e_i\slashed{p}_i+M),\tilde{H}^{\r}\right]_i\right]_i\partial_{\n}F_{\a\m}(\s v_i) W_{v_2}W_{v_1}\right\rangle}{\langle W_{v_2}W_{v_1}\rangle}  \right]
\nn \\
&&+\sum_{i=1}^2\tilde{H}^{\r} \int d^3\s \frac{e^2 v_i^2 v_i^{\a}v_i^{\b}\eta^{\m\n}}{4p_i\cdot v_i} \frac{\left\langle F_{\a\m}(v_i\s_1)F_{\b\n}(v_i\s_2) W_{v_1}W_{v_2}\right\rangle}{\langle W_{v_1}W_{v_2}\rangle}\nn \\ 
&&\left.+\sum_{i=1}^2\int d^2\s\frac{e_iev_i^2v^\a_i}{2p_i\cdot v_i}\frac{\left\langle \partial_{\n}F_{\a\m}(v_i\s) W_{v_1}W_{v_2}\right\rangle}{\langle W_{v_1}W_{v_2}\rangle} \left[\g^{\m}\g^{\n},\tilde{H}^{\r}\right]_i\right),\label{eq:final-result}
\eea
where we have introduced a Dirac matrix ordering operator $[x,y]_i$ which is $xy$ when $i=1$ and is $yx$ when $i=2$.

The factorization formula in Eq.\ (\ref{eq:final-result}) is to be understood as a recursive definition for the hard function $\tilde{H}$, which is clearly distinct from the hard function in Eq.\ (\ref{eq:heavy-quark-LP}). Here, we have separated the leading-power hard function into its NLP low-energy component and a new hard function $\tilde{H}$. Let us make some  comments about the factorization at NLP in Eq.\ (\ref{eq:final-result}). Unlike at leading power, the factorization is not a product of a hard function and a single soft function. Instead, we find that the soft function consists of four components, each with a nontrivial insertion of field-strength tensors integrated along the worldline of each particle.  The hard function also contracts its Dirac indices into the soft function. Finally, there is a Low’s theorem-like term, which also acts as a derivative operator on the hard function. This factorization is significantly more intricate than the leading power factorization theorems. 

The first line of Eq.\ (\ref{eq:final-result}) consists of a Low’s theorem-like term, the second line is the hard vertex correction, and the third and fourth lines are the double and single G-photon insertions, respectively. In QED, a further separation of the hard function into NLP terms and a new hard function is somewhat artificial. However, such a separation in the non-Abelian theory is expected to separate out perturbative and nonperturbative modes at next-to-leading power.

We would now like to show that the relationship in Eq.\ (\ref{eq:final-result}) is valid to all orders in perturbation theory. One immediate question that arises is whether at three and higher loops, a new term contributes to the NLP soft function involving more G-photons.  We will argue in the next section that this does not occur.
 This would run counter to the Wilsonian philosophy of organizing physics by scale. One does not expect that infinite terms appear at a particular order in the power expansion.

 \section{Higher loops and generalizations}
 \label{sec:all-order-analysis}
 In this section, we provide brief arguments supporting the assertion that the formula in Eq.\ (\ref{eq:final-result})  holds to all orders—both with light fermions (when the soft functions are nontrivial) and without light fermions, in which case the soft functions are two-loop exact, provided that the anomalous magnetic moment correction is computed to the desired loop order.
 
  As an application of our formalism, we derive the NLP soft photon theorem in the context of soft photon production in association with a heavy fermion.
   Such a soft photon theorem is not one-loop exact, even in the absence of massless fermions, due to the anomalous magnetic moment. In the presence of massless fermions, the matrix elements themselves are nontrivial to all orders in perturbation theory. When a real soft photon is observed in the final state (as opposed to the virtual soft photons considered so far), the separation into leading and next-to-leading power becomes a systematic expansion of the soft photon amplitude in powers of $\o_k$, the energy of the soft photon. 
  
  To argue that the result in Eq.\ (\ref{eq:final-result}) holds to all orders, we enumerate all possible corrections from higher-order graphs.
  \begin{itemize}
  	\item Three or more G-photons attaching to the fermion line.
  	\item Multiple hard subdiagrams and the resulting composite vertex insertions near a pinch surface.
  	\item Pinch surfaces with multiple soft photons attaching to the hard subgraphs.
  	\item Interference between multiple K-photons and the G-photons of interest.
  \end{itemize}
  
  We argue in turn that each potential obstruction to factorization, as in Eq.\ (\ref{eq:final-result}), is either power-suppressed—as in the first three cases—or absent entirely due to generalizations of the Ward identities in Eqs.\ (\ref{eq:ward-id}), (\ref{eq:ward-id-2}), and their generalizations. 
  
  Let us first show that there is no interference between K and G photons. 
  
  Consider an arbitrary soft pinch of the form shown in Fig.\ \ref{fig:pinches}. We adopt the strategy of decomposing the fermion line into its individual components.
  \bea
  \Gamma^{\r}=\int_{l_i,k_j}F_{n}^{\m_1\ldots \m_n}(l_1,\ldots l_n,p_1)\tilde{H}^{\r}F_{n}^{\n_1\ldots \n_m}(k_1,\ldots k_m,p_2)S_{\m_1\ldots\m_n\n_1 \ldots \n_m}(\{l_i,k_j\}),
  \eea
  where all topologies internal to the soft subgraph have been summed over in  $S$, including momentum-conserving delta functions for connected subgraphs. For example, at one loop $S(l,k)$ is 
  \bea
  S_{\m\n}^{(1)}(l,k)=\frac{-ie^2\eta_{\m\n}}{l^2+i\e}(2\pi)^d\d^d(l-k), 
  \eea
  and at two loops, it is equal to 
  \bea
  S_{\m_1\m_2\n_1\n_2}^{(2)}(l_1,l_2,k_1,k_2)=\frac{1}{2}\frac{-ie^2\eta_{\m_1\n_1}}{l_1^2+i\e}\frac{-ie^2\eta_{\m_2\n_2}}{l_2^2+i\e}\left((2\pi)^{2d}\d^d(l_l-k_1)\d^d(l_2-k_2)\right). 
  \eea
  We observe that the sum over different attachments to the fermion line is carried out in the fermion line Dirac matrix $F$. This form of $S$ repeats to all orders in perturbation theory when no light fermions are present in the IR. However, $S$ is more complicated, starting from four loops in the presence of light fermions.
  
  Returning to the fermion line Dirac matrix element, we write it as
  \bea
F_{n}^{\m_1\ldots \m_n}(l_1,\ldots l_n,p_1)&=&\left( K^{\m_1}_{\,\a_1}(l_1,p_1) +  G^{\m_1}_{\,\a_1}(l_i,p_1)
\right)\ldots \left( K^{\m_n}_{\,\a_n}(l_n,p_1) +  G^{\m_n}_{\,\a_n}(l_n,p_1)
\right) \nn \\&& \times F_{n}^{\a_1\ldots \a_n}(l_1,\ldots l_n,p_1)\nn \\ 
&& \hspace{-2cm}=\sum \limits_{j=0}^n \sum_{(a_1,a_2\ldots a_j)}G^{\m_{a_1}}_{\,\a_{a_1}}\ldots G^{\m_{a_k}}_{\,\a_{a_k}}
K^{\m_{b_1}}_{\,\a_{b_1}}\ldots K^{\m_{b_{n-k}}}_{\,\a_{b_{n-k}}}F_{n}^{\a_1\ldots \a_n}(l_1,\ldots l_n,p_1),\label{eq:fermion-line-decomp-general}
  \eea
  where we have introduced the $(b_1\ldots b_{n-k})=\{b_i:b_{i}\in [1,n], b_{i}\not\in \vec{a} \}$. 
  Let us write an explicit expression for the fermion line
  \bea
  F_{n}^{\m_1\ldots \m_n}(l_1,\ldots l_n,p_1)&=& \sum_{\pi_{a}(n)}
  \left(\prod\limits_{j=1}^{n}\g^{\m_{\pi^j_a}}\frac{\slashed{p}_1-\slashed{L}_{j}+M}{(p_1-L_j)^2-M^2+i\e}\right), \nn \\
  L_j&=&\sum_{i=1}^{j} l_{\pi_a^i}, \label{eq:fermion-line-general}
  \eea
  where $a\in [1,n!]$ labels a specific permutation of $n$ elements, and $\pi^{j}_a$ labels the $j$th element of the permutation $a$. In the matrix product
 displayed in Eq.\ (\ref{eq:fermion-line-general}), the product is expanded from left to right. 
 
 As in Secs.\ \ref{sec:one-loop-analysis} and \ref{sec:two-loop-analysis}, the K-photons satisfy a Ward identity:
 \bea
  K^{\a_m}_{\;\m_m}(l_m,p_1)F_{n}^{\m_1\ldots \m_n}(l_1,\ldots l_n,p_1)=\frac{-p_1^{\a_m}}{p_1 \cdot l_m}F_{n-1}^{\m_1\ldots\bar{\m}_m \ldots \m_n}(l_1,\ldots \bar{l}_m \ldots l_n,p_1),\label{eq:ward-identity-general}
 \eea 
 where $\bar{l}_m$ means $l_m$ does not appear in the list. 
 We can use the Ward identity in Eq.\ (\ref{eq:ward-identity-general}) to write
 \bea
 G^{\m_{a_1}}_{\,\a_{a_1}}\ldots G^{\m_{a_k}}_{\,\a_{a_k}}
 K^{\m_{b_1}}_{\,\a_{b_1}}\ldots K^{\m_{b_{n-k}}}_{\,\a_{b_{n-k}}}F_{n}^{\a_1\ldots \a_n}(l_1,\ldots l_n,p_1)&=&\nn \\ &&\hspace{-6cm}\left(\prod_{i=1}^{n-k}\frac{-p_1^{\a_{b_i}}}{p_1 \cdot l_{b_i}}\right) G^{\m_{a_1}}_{\,\a_{a_1}}\ldots G^{\m_{a_k}}_{\,\a_{a_k}}F_{k}^{\a_{a_1}\ldots \a_{a_k}}(l_{a_1},\ldots l_{a_k},p_1).
 \eea
  
 This identity demonstrates that K-photons can be peeled off recursively, leaving behind a lower-order graph with G-photons. We conclude that all K-photons decouple from G-photons.
  
  Let us now briefly argue that terms with three or more G-photon insertions in Eq.\ (\ref{eq:fermion-line-decomp-general}) are power-suppressed relative to the one and two G-photon terms.
  
 Consider the fermion line with three-photon attachments:
  \bea
  \bar{u}(p_1)F_3^{\m_1\m_2\m_3}(l_1,l_2,l_3)&= &\bar{u}(p_1)\g^{\m_1} \frac{\slashed{p}_1-\slashed{l}_1+M}{(p_1-l_1)^2-M^2+i\e}\g^{\m_2} \frac{\slashed{p}_1-\slashed{l}_{12}+M}{(p_1-l_{12})^2-M^2+i\e} \nn \\ 
  &&\times \g^{\m_3} \frac{\slashed{p}_1-\slashed{l}_{123}+M}{(p_1-l_{123})^2-M^2+i\e}+\text{permutations}.
  \eea
  When studying the triple G-photon insertion, $G^{\m_1}_{\;\a_1}(l_1),G^{\m_2}_{\;\a_2}(l_2)G^{\m_3}_{\;\a_3}(l_3)F^{\a_1\a_2\a_3}$, we observe that at NLP, we can retain at most one factor of soft momentum $\slashed{l}_i$, since retaining two or more powers of soft momentum in the numerator is doubly power suppressed. Thus, we may decompose the NLP fermion line as
  \bea
  G^{\m_1}_{\;\a_1}(l_1),G^{\m_2}_{\;\a_2}(l_2)G^{\m_3}_{\;\a_3}(l_3)F^{\a_1\a_2\a_3}_{3,NLP}&=& G^{\m_1}_{\;\a_1}(l_1),G^{\m_2}_{\;\a_2}(l_2)G^{\m_3}_{\;\a_3}(l_3)F^{\a_1\a_2\a_3}_{3,\text{NLP},a}
  \nn \\ &&\hspace{-6cm}+G^{\m_1}_{\;\a_1}(l_1),G^{\m_2}_{\;\a_2}(l_2)G^{\m_3}_{\;\a_3}(l_3)F^{\a_1\a_2\a_3}_{3,\text{NLP},b}+G^{\m_1}_{\;\a_1}(l_1),G^{\m_2}_{\;\a_2}(l_2)G^{\m_3}_{\;\a_3}(l_3)F^{\a_1\a_2\a_3}_{3,\text{NLP},c}.
  \eea
  In the term  $G^{\m_1}_{\;\a_1}(l_1),G^{\m_2}_{\;\a_2}(l_2)G^{\m_3}_{\;\a_3}(l_3)F^{\a_1\a_2\a_3}_{3,\text{NLP},a}$, if the numerator of the left most fermion line is replaced by the soft momentum, we obtain
  \bea
  \bar{u}(p_1)G^{\m_1}_{\;\a_1}(l_1),G^{\m_2}_{\;\a_2}(l_2)G^{\m_3}_{\;\a_3}(l_3)F^{\a_1\a_2\a_3}_{3,\text{NLP},a}&= &\bar{u}(p_1)\slashed{G}^{\m_1} \frac{-\slashed{l}_1}{(p_1-l)^2-M^2+i\e} \label{eq:G-triple-photon} \\&&\hspace{-5cm} \times \slashed{G}^{\m_2} \frac{\slashed{p}_1+M}{(p_1-l_{12})^2-M^2+i\e}  \slashed{G}^{\m_3} \frac{\slashed{p}_1+M}{(p_1-l_{123})^2-M^2+i\e}+\text{permutations},\nn 
  \eea
  where we have introduced $\slashed{G}^{\m}=G^{\m}_{\a}\g^{\a}$. It is straightforward to verify that the expression in Eq.\ (\ref{eq:G-triple-photon}) vanishes identically using
  \bea
  \slashed{G}^{\m_3}(\slashed{p}_1+M)&=&(-\slashed{p}_1+M)  \slashed{G}^{\m_3} \nn \\ 
  (\slashed{p}_1+M)(-\slashed{p}_1+M)&=&0. \label{eq:G-photon-identity}
  \eea
  Similarly, the term labeled $G^{\m_1}_{\;\a_1}(l_1),G^{\m_2}_{\;\a_2}(l_2)G^{\m_3}_{\;\a_3}(l_3)F^{\a_1\a_2\a_3}_{3,\text{NLP},b}$, in which the numerator of the middle fermion line is replaced by the soft momentum, vanishes due to the identity in the first line of  Eq.\ (\ref{eq:G-photon-identity}) and the Dirac equation on the external spinor. Likewise, the term with the right-most fermion line, when replaced by the soft momentum,
  $G^{\m_1}_{\;\a_1}(l_1),G^{\m_2}_{\;\a_2}(l_2)G^{\m_3}_{\;\a_3}(l_3)F^{\a_1\a_2\a_3}_{3,\text{NLP},c}$, also vanishes. This argument generalizes to any number (greater than or equal to three) of G-photons. As a result, terms with more than three G-photons are power-suppressed.
  
\begin{figure}
  	\begin{center}
  		\includegraphics[width=0.5\textwidth]{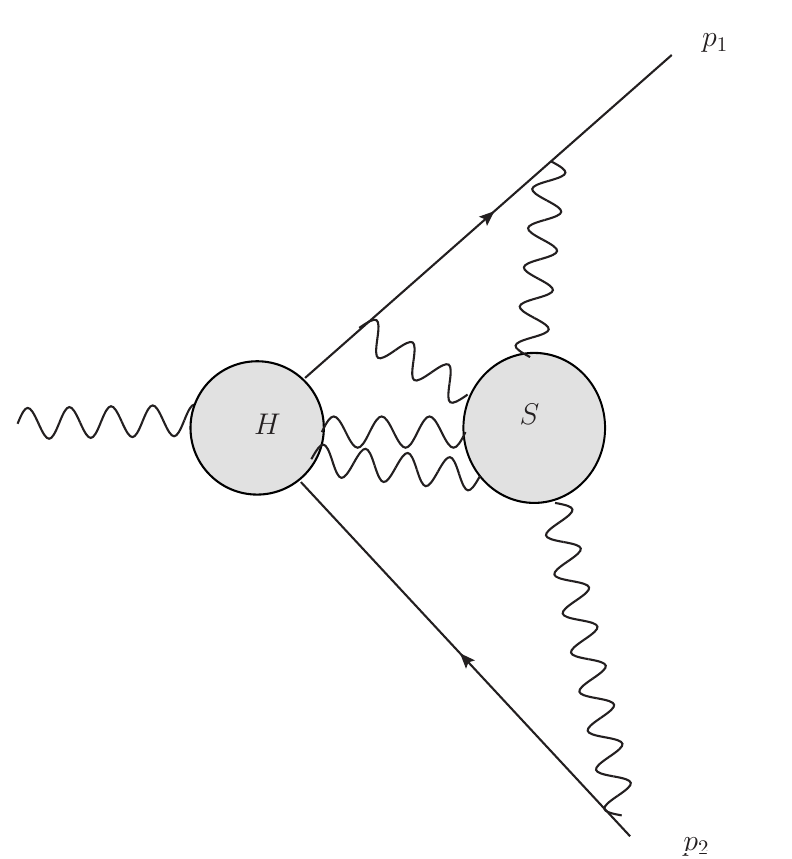}
  		\caption{A double power suppressed pinch surface with two soft photons entering the hard part.}
  		\label{fig:multiple-sh-pinch}
  	\end{center}
  \end{figure}

  In on-shell renormalization,  composite hard vertices embedded in the pinched subdiagram always begin at next-to-leading power (at $O(l)$, where $l$ is the soft momentum). As a result, at most one composite vertex is inserted at NLP. Finally, it is easy to see that pinches of the type shown in Fig.\ \ref{fig:multiple-sh-pinch} are doubly power suppressed because at least two eikonal denominators are missing in comparison to the leading power expression. We conclude that the result in Eq.\ (\ref{eq:final-result}) is expected to hold to all orders in perturbation theory. 
   
  We now turn to the soft photon theorem  at NLP. Before we proceed, let us briefly review the leading power result of \cite{Ma:2023gir}. At leading power, the soft photon theorem is
  \bea
  \mathcal{M}_3^{\r\m}(p_1,p_2,k)\e_{\m}(k)&=&\left(\sum_{i=1}^2\frac{e_ip_i\cdot \e}{p_i\cdot k}+\frac{\langle \e^{\m}j_\m(k)W_{v_1}W_{v_2}\rangle}{\langle W_{v_1} W_{v_2}\rangle}\right)\Gamma^\r(p_1,p_2)+O(\o_k^0).
  \eea
  The first term is the K-photon attachment to the fermion line, while the second term is to be interpreted as the insertion of the electromagnetic current in the leading power soft matrix element.  We now would like to compute the first correction to this formula, i.e., the $O(\o_k^0)$ term. We label the  $O(\o_k^0)$ by $\mathcal{M}_{3,\text{NLP}}^{\r\m}(p_1,p_2,k)\e_{\m}(k)$. 
  
  As before, we expect four terms: a single real G photon insertion, one real and one virtual G photon insertion, composite vertex corrections to the real G photon, and Low's like radiation of the G-photon from the hard part.  Alternatively, one could radiate a soft photon from an NLP soft matrix element, which will automatically yield NLP soft photon radiation. However, radiation from an NLP soft matrix element requires virtual soft fermions. In the absence of light fermions, such matrix elements yield zero. Therefore, it is natural to make the decomposition
  \bea
  \mathcal{M}_{3,\text{NLP}}^{\r\m}(p_1,p_2,k)\e_{\m}(k)=V_{3}^{\r\m}(p_1,p_2,k)\e_{\m}(k)+R_{3}^{\r\m}(p_1,p_2,k)\e_{\m}(k),
  \eea
  where $V$ only produces NLP soft photons through virtual soft fermions, and $R$ corresponds to NLP corrections to the attachment of the soft photon to fermion lines. It is in the calculation of $V_3$ that the formula in Eq.\ (\ref{eq:final-result}) comes to life.
  \bea
  V_{3}^{\r\m}(p_1,p_2,k)\e_{\m}(k)&=&\nn \\ && \hspace{-3cm} \left.\langle W_{v_2}W_{v_1}\rangle \sum_{i=1}^2\frac{\partial {H}^{\rho}(p_1,p_2)}{\partial p_i^{\b}}\int d\s \, {e_iev_i^{\a}}\frac{\langle \e^{\m}j_\m(k) F_{\b\a}(\s v_i) W_{v_2}W_{v_1}\rangle}{\langle W_{v_2}W_{v_1}\rangle}\right. \nn \\ 
  &&\hspace{-3cm}+\sum_{i=1}^2F_2(0) \left[\int d^2\s \frac{ie_iev_i^2v_i^{\a}}{M p_i\cdot v_i}\frac{\left\langle \left[\s^{\m\n},\left[i(e_i\slashed{p}_i+M),\tilde{H}^{\r}\right]_i\right]_i\partial_{\n}F_{\a\m}(\s v_i) W_{v_2}W_{v_1}\e^{\m}j_\m(k)\right\rangle}{\langle W_{v_2}W_{v_1}\rangle} \right]
  \nn \\
  &&\hspace{-3cm}+\sum_{i=1}^2\Gamma^{\r} \int d^3\s \frac{e^2 v_i^2 v_i^{\a}v_i^{\b}\eta^{\m\n}}{4p_i\cdot v_i} \frac{\left\langle F_{\a\m}(v_i\s_1)F_{\b\n}(v_i\s_2) W_{v_1}W_{v_2}\e^{\m}j_\m(k)\right\rangle}{\langle W_{v_1}W_{v_2}\rangle}\nn \\ 
  &&\hspace{-3cm}\left.+\sum_{i=1}^2\int d^2\s\frac{e_iev_i^2v^\a_i}{2p_i\cdot v_i}\frac{\left\langle \partial_{\n}F_{\a\m}(v_i\s) W_{v_1}W_{v_2}\e^{\m}j_\m(k) \right\rangle}{\langle W_{v_1}W_{v_2}\rangle} \left[\g^{\m}\g^{\n},\Gamma^{\r}\right]_i\right),
  \eea 
  It is easy to read off $R_3$, using the analyses in Sec.\ \ref{sec:one-loop-analysis} and \ref{sec:two-loop-analysis}
  \bea
  R_{3}^{\r\m}(p_1,p_2,k)\e_{\m}(k)&=& \langle W_{v_2}W_{v_1}\rangle \sum_{i=1}^2\frac{\partial {H}^{\rho}(p_1,p_2)}{\partial p_i^{\b}} \left(k_\b \e_{\a}-k_\a \e_{\b}\right)\left(\frac{e_iep_i^\a}{p_{i}\cdot k}\right)\nn \\ 
 && \hspace{-3cm}+\sum_{i=1}^2F_2(0)  \frac{ie_ie p_i^{\a}}{M (p_i\cdot k)^2}k_{\n}(k_\a\e_{\m}-\e_{\a}k_{\m}) \left[\s^{\m\n},\left[i(e_i\slashed{p}_i+M),\Gamma^{\r}\right]_i\right]_i\nn \\ 
 && \hspace{-3cm}+\sum_{i=1}^2\frac{e_ie p_i^{\a}}{2 (p_i\cdot k)^2}k_{\n}(k_\a\e_{\m}-\e_{\a}k_{\m})\left[\g^{\m}\g^{\n},\Gamma^{\r}\right]_i\nn\\ &&
 \hspace{-3cm} +\sum_{i=1}^2\Gamma^{\r} \int _l \frac{e^2  p_i^{\a}p_i^{\b}\eta^{\m\n}(k_\a\e_{\mu}-\e_{\a}k_{\m})(l_{\b}\d_\n^\g-l_\n\d_\b^\g)}{4(p_i\cdot k)(p_i\cdot l)(p_i\cdot(k+l))l^2} \frac{\left\langle \tilde{A}_{\g}(l) W_{v_1}W_{v_2}\right\rangle}{\langle W_{v_1}W_{v_2}\rangle}
  \eea
  Where the first term is the standard Low's theorem, and all but the second term can be evaluated using a one-loop computation if there are no massless fermions.  In particular, the third and fourth terms correspond to the radiated photon being G-polarized and the radiated photon, along with one virtual photon, being G-polarized, respectively.  
  The coupling of the radiated G photon to the anomalous magnetic moment receives corrections to all orders.  Finally, let us say some words about self-energy graph. In the on-shell scheme we have adopted throughout this paper, at leading power, any self-energy on the external leg can be set to zero. However, when accounting for the individual K, G components at the level of the integrand, we have observed that they need to be included in a scheme-agnostic fashion. Therefore, one may first include all self energies in a scheme agnostic fashion, and finally choose counterterms relevant to the on-shell scheme.
  
  Let us conclude this section by summarizing what has been achieved. We have argued that power counting suggests that the formula in Eq.\ (\ref{eq:final-result}) is expected to hold to all orders. We have also used the foregoing analysis to read off the NLP soft photon theorem, which can be done  straightforwardly.  
 \section{Conclusions}   
\label{sec:conclusions}
In this work, we have studied the question of power corrections to factorization theorems in the abelian gauge theory. We have shown that the heavy electron form-factor factorizes at next-to-leading power to all orders in perturbation theory. We have also shown that there are three sources of power corrections: a single soft photon becoming G polarized, two soft photons becoming G polarized, and Taylor expansion of the hard part in soft momenta. Each of these corrections corresponded to a matrix element coupled to the hard function in a gauge invariant fashion (with insertion of field strength tensors and their derivatives). 

Let us make some remarks about the class of all power corrections to QED amplitudes. Since all higher power corrections come from soft (G) photons coupling to the fermion line, the infrared dynamics of QED is completely specified by Wilson lines and field strength insertions.  Also of importance for computing the hard functions for higher power corrections are higher derivatives of hard functions (where fermion propagators occur with increasing exponents). An IR effective field theory of massive QED, therefore, consists of three qualitative ingredients: Wilson lines, field strength tensors, and massless matter. This is the expectation one might have at the outset. Factorization theorems enable us to infer which combination of field-strength, Wilson line insertion correlators are of observable importance from the perspective of high-energy scattering.     

We conjecture that this factorization holds in the non-abelian theory and leave the proof to future work. One question that remains open is that of the generalization of NLP soft functions to the non-abelian theory. Does the non-abelian theory admit power correction when a single G-gluon attaches to the fermion line? Can we consistently peel off K-gluons without affecting the color structure of the G-gluons? The final goal of such a program might be to prove that the inclusive production of quarkonium factorizes at NLP.  

Yet another set of questions involves resummation in the non-abelian theory. Real soft radiation in the non-abelian theory can be observed. In such an observable, can we resum the large logarithms at next-to-leading power. Such a resummation program will be of importance to future colliders. 
\section*{Acknowledgements}
I would like to thank Eric Laenen for a careful reading of the manuscript. I would also like to thank George Sterman for collaboration on related problems.

\end{document}